\documentclass[5p]{elsarticle}

\usepackage{lineno,hyperref}
\modulolinenumbers[50000]
\usepackage{amssymb}
\usepackage{mathrsfs}
\usepackage{amsmath}
\usepackage{amsthm}
\usepackage{multirow}
\usepackage{array}
\usepackage{tikz}
\usepackage[labelfont=bf]{caption}
\usepackage[font=footnotesize]{subcaption}
\usepackage{stfloats}
\usepackage[noline,ruled,commentsnumbered,linesnumbered,titlenumbered]{algorithm2e}

\newcolumntype{x}[1]{>{\centering\arraybackslash\hspace{0pt}}m{#1}}
\usetikzlibrary{shapes.geometric,arrows}
\usetikzlibrary{positioning,decorations.markings}
\tikzstyle{startstop} = [rectangle, rounded corners, minimum width=1cm,
                         minimum height=0.5cm,text centered, text width=1cm,
                         draw=black, fill=gray!20]
\tikzstyle{init} = [rectangle, rounded corners, minimum width=2cm,
                         minimum height=1cm,text centered, text width=2cm,
                         draw=black, fill=gray!20]
\tikzstyle{io} = [trapezium, trapezium left angle=70,
                  trapezium right angle=110, minimum width=3cm,
                  minimum height=1cm, text centered, draw=black, fill=gray!20]
\tikzstyle{process} = [rectangle, minimum width=2cm, minimum height=1cm,
                       text centered, text width=3.5cm, draw=black,
                       fill=gray!20, inner xsep=-1.0pt]
\tikzstyle{decision} = [diamond, minimum width=3cm, minimum height=1cm,
                        text centered, text width=4.2cm, draw=black,
                        fill=gray!20, inner sep=-5pt]
\tikzstyle{dot} = [circle, minimum size=0.5cm, draw=black, fill=gray!20]
\tikzstyle{input} = [coordinate]
\tikzstyle{arrow}=[line width=0.2mm,draw=gray,-triangle 45,postaction={draw, line width=0.3mm, shorten >=0.3mm, -}]

\journal{Energy and Buildings}
\begin{document}

\begin{frontmatter}

  \title{Automatic HVAC Control with Real-time Occupancy Recognition and
  Simulation-guided Model Predictive Control in Low-cost Embedded System}

  \author{Muhammad Aftab, Chien Chen, Chi-Kin Chau and Talal Rahwan\tnoteref{myfootnote}}
  \address{Masdar Institute}
  \tnotetext[myfootnote]{Email: \{ckchau, trahwan\}@masdar.ac.ae}

  \begin{abstract}
    Intelligent building automation systems can reduce the energy consumption of heating, ventilation and air-conditioning (HVAC) units by sensing the comfort requirements automatically and scheduling the HVAC operations dynamically. Traditional building automation systems rely on fairly inaccurate occupancy sensors and basic predictive control using oversimplified building thermal response models, all of which prevent such systems from reaching their full potential. Such limitations can now be avoided due to the recent developments in embedded system technologies, which provide viable low-cost computing platforms with powerful processors and sizeable memory storage in a small footprint. As a result, building automation systems can now efficiently execute highly-sophisticated computational tasks, such as real-time video processing and accurate thermal-response simulations. With this in mind, we designed and implemented an occupancy-predictive HVAC control system in a low-cost yet powerful embedded system (using Raspberry Pi 3) to demonstrate the following key features for building automation: (1) real-time occupancy recognition using video-processing and machine-learning techniques, (2) dynamic analysis and prediction of occupancy patterns, and (3) model predictive control for HVAC operations guided by real-time building thermal response simulations (using an on-board EnergyPlus simulator). We deployed and evaluated our system for providing automatic HVAC control in the large public indoor space of a mosque, thereby achieving significant energy savings.

  \end{abstract}


\end{frontmatter}

\section{Introduction} \label{sec:intro}

\noindent Heating, ventilation, and air-conditioning (HVAC) units, which are a primary target of building automation, make up almost 50\% of the energy consumed in both residential and commercial buildings \cite{HVACConsumption}.  In general, building automation systems aim to intelligently control building facilities in response to dynamic environmental factors, while maintaining satisfactory performance in energy consumption and comfort. The primary functions of a building automation system include: (1) sensing of the environmental factors by measurements, and (2) optimizing control strategies based on the current and predictive states of building and occupancy. These tasks require an integrated process of sensing, computation, and control.

Traditional building automation systems rely on fairly inaccurate occupancy sensors, which hinder the responsiveness of automation systems. For example, passive infrared and ultra-sound occupancy sensors produce poor accuracy, because they are unable to determine the occupancy state adequately when occupants remain stationary for a prolonged period of time. They also have a limited range which hinders their performance, especially in a large area. More accurate sensing technology, such as cameras that use visible or infra-red lights, can significantly improve the accuracy of occupancy recognition.

On the other hand, model predictive control, by which the future thermal response and external environmental factors are anticipated to make control decisions accordingly, has been considered in a number of studies \cite{mpc1,mpc2,mpc3,mpc5,mpc6,mpc7} which are shown to be more effective than classical PID and hysteresis controllers that do not consider anticipated events. However, these studies are often based on time-invariant, first-principle linear models (also known as lumped element resistance-capacitance (RC) models \cite{lti_rc}), considering only simple building geometry and single-zone in near-future time horizon. Although these linear models are easier for calibration (e.g., using frequency domain decomposition, or subspace system identification methods \cite{lti_rc, mpc_rc_calib1, mpc_rc_calib2}), the error accumulates considerably when a longer time horizon is considered in model predictive control. While non-linear models are rather complicated and impractical, other alternatives based on physical models of building thermal response can provide a feasible solution.

Recently, there have been remarkable advances in embedded system technologies, which provide low-cost platforms with powerful processors and sizeable memory storage in a small footprint. In particular, the emergence of \emph{system-on-a-chip} technology \cite{soc}, which integrates all major components of a computer into a single chip, can provide versatile computing platforms with low-power consumption and mobile network connectivity in a cost-effect\-ive manner for mass production. As a result, smartphones are able to rapidly evolve from single-core to multi-core processors with a low incremental production cost. Notably, the \emph{Raspberry Pi} project \cite{rpi}, which originally aimed to provide affordable solutions for the teaching of computer science, has rapidly evolved for a wide range of advanced scientific projects. Therefore, there are plenty of opportunities to harness recent embedded system technologies in intelligent building automation systems. Particularly, sophisticated computational tasks can be conducted on these embedded systems efficiently, such as real-time video processing and accurate building thermal response simulation (e.g., \cite{embedded_ref2}).

\medskip

With this in mind, we designed and implemented an occupancy-predictive HVAC control system in a low-cost yet powerful embedded system (using Raspberry Pi 3) for building automation. The rest of the paper is organized as follows. In Section~\ref{sec:related}, we present the background information and literature review. In the remaining sections, we highlight three key features of our system.

\paragraph{{\em(Section~\ref{sec:recognition})} Real-time Video-based Occupancy Recognition}
We apply advanced video-processing techniques to analyze the features of occupants from video cameras, and automatically classify and infer the states of occupancy. Moreover, we consider privacy enhancement using a frosted lens. Our system achieves 80-90\% accuracy for occupancy recognition by real-time video processing. Furthermore, we improve the performance of our occupancy recognition by using Machine Learning for considerably crowded settings.

\paragraph{{\em(Section~\ref{sec:prediction})} Dynamic Occupancy Prediction}
We employ various linear and non-linear regression models to capture and predict occupancy trends according to different day-of-week, seasonal patterns, etc.  We present general as well as domain-specific approaches for occupancy prediction. Our models are able to identify the future occupancy trends considering a variety of dynamic usage patterns.

\paragraph{{\em(Sections~\ref{sec:simulation}-\ref{sec:eval})} Simulation-guided Model Predictive Control}
We employ \emph{EnergyPlus} simulator \cite{eplus} for real-time HVAC control. We ported EnergyPlus simulator to the Raspberry Pi embedded system platform for simulation-guided model predictive control. A co-simulation framework is utilized to provide accurate building thermal response simulation under proper calibration. Noteworthily, we also release our Raspberry Pi version of EnergyPlus publicly \cite{eplus_rpi} to enable other researchers to take advantage of our work for future building automation projects.

\medskip

Our automatic HVAC control system is intended for public indoor spaces, such as corridors, libraries, or communal areas. Unlike private spaces such as homes, these public indoor spaces are not controlled by a particular occupant and can be affected by a diverse set of occupancy patterns.  Such patterns tend to vary more dynamically in public spaces compared to private ones, posing challenges for effective occupancy sensing and prediction systems.

\begin{figure}[htp!]
  \centering
  \includegraphics[width=1.0\linewidth]{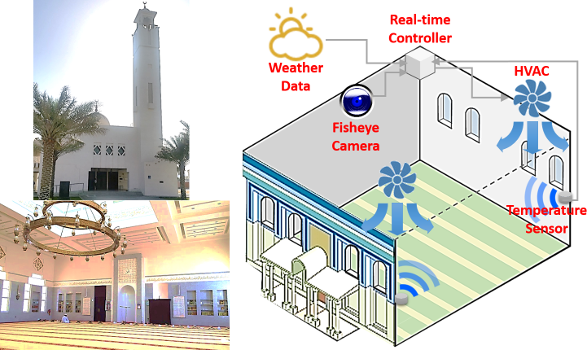}
  \caption{A large public indoor space of a mosque is used as a testbed for our automatic HVAC control system. Fisheye video camera, temperature and humidity sensors, as well as real-time controller for HVAC have been deployed in our testbed.}
  \label{fig:mosque}
\end{figure}

In particular, our system is deployed and evaluated for providing automatic HVAC control in the large public indoor space of a mosque (see Figure~\ref{fig:mosque}), which is the worship place for followers of Islam. Typically, mosques have large public spaces, and are open 24-hours a day and 7-days a week. There are nearly 5,000 mosques in the UAE \cite{NumberOfMusquesInUAE}, and over 55,000 mosques in Saudi Arabia \cite{NumberOfMusquesInSaudi}. Due to the hot climate in this region, HVAC is required on a regular basis. The results obtained from our testbed implementation in Section~\ref{sec:testbed} demonstrate the significant energy savings that can be achieved by using automatic HVAC control systems in public indoor spaces.

\section{Background and Related Work}\label{sec:related}

\noindent In this section, we present the background and related work of our system. In particular, this section consists of two subsections. The first provides a brief review of occupancy recognition and prediction. The second subsection compares three main approaches of model predictive HVAC control.

\subsection{Occupancy Recognition and Prediction} 

\noindent Any object with a temperature higher than perfect zero emits heat in the form of radiation. The conventional passive infra-red (PIR) sensors can be used to detect a certain wavelength when a person is near the sensor.  Previous papers  \cite{pir1,pir2,pir3} demonstrated the possibility of applying this to control lighting systems. However, the disadvantage of the PIR sensor becomes evident when the occupant remains stationary for a certain period of time. In particular, PIR sensors are designed to detect changes in the movement, and if a person remains stationary in front of the sensor, then as far as the sensor is concerned, there will be no change in the movement, leading to an erroneous observation.

Video-based occupancy-detection algorithms can provide better accuracy than their PIR counterparts. One such algorithm was proposed in \cite{camera1}, whereby certain features (such as edges, textures, etc.) are extracted from the video and used in a regression model to estimate the number of occupants. Unfortunately, this algorithm is computationally intensive, rendering it unsuitable for implementations on embedded systems such as Raspberry Pi. In contrast, our algorithm can detect occupancy in real-time, even when implemented on an embedded system, as is the case with our testbed. An alternative algorithm was proposed in \cite{camera2}. Although this algorithm is computationally less intensive than the previous one, it nevertheless relies heavily on identifying the heads of the occupants. This head-detection process is particularly challenging in our testbed, as it is common for a typical occupant to have a head cover indistinguishable from his or her outfit. Since our system does not require head detection, it is insensitive to whether or not the occupant is wearing a head cover.

In addition to occupancy recognition, occupancy prediction is required to anticipate building usage and control pre-cooling in advance. A variety of techniques to predict occupancy have been proposed in the literature, including statistical analysis, machine learning, and stochastic modeling. A comprehensive review of  occupancy prediction techniques is provided in \cite{review_pred1, review_pred2}.

\subsection{Model Predictive Control}
\noindent There are three major approaches to model predictive control in the literature:

\begin{itemize}

\item{\em LTI Model Predictive Control}: This uses a linear time-invariant (LTI) mathematical model, which is a simplified thermal dynamics model considering only a near-future short time horizon. A common approach is to use an RC model to capture the first-order heat transfer dynamics. It is suitable for simple settings, such as a single zone with simple building geometry. There are usually a small number of parameters in the model. LTI model predictive control is explored in \cite{lti_rc,mpc_rc_calib1,mpc_rc_calib2,dong2011-1,dong2014}.

\item{\em Non-linear Model Predictive Control}: Many real-world systems exhibit rich non-linearity. There are many general non-linear mathematical models of system dynamics, such as Volterra series, neural networks and NARMAX models. However, most non-linear models require a large parameter space, which is difficult to calibrate from measurements. The use of non-linear models for predictive control is investigated in \cite{neuralnetworks,narmax}.

\item{\em  Simulation-guided Model Predictive Control}: In this approach, model predictive control is guided by real-time physical model simulators considering future anticipated events. For buildings, there are a number of simulators, such as \emph{EnergyPlus} and \emph{TRNSYS}, that are much more accurate than LTI models, and also easier to calibrate than general non-linear models. Reviews of different building simulators and their merits are presented in \cite{mpc5,simulators}.

\end{itemize}

For the above control approaches to be effective, it is crucial to calibrate the model parameters so that the model response is consistent with the empirical data. Several model calibration methods have been proposed in the literature. These methods can be broadly categorized as {\em manual} or {\em automated}.  In particular, {\em manual} approaches require the modeler to intervene repeatedly and make adjustments, whereas {\em automated} approaches use mathematical and statistical models to automate the calibration process. A review of model calibration can be found in \cite{calib_review}.

Several previous studies used simulation programs to facilitate model predictive control. Specifically, in \cite{mpc5,mpc7,cosim_mpc4}, the authors employed co-simulation for model predictive control (MPC) with EnergyPlus. However, these studies did not implement the MPC models in real-world HVAC systems and were limited to simulations. Other studies \cite{sim_based_mpc,cosim_mpc5} tested the MPC models with real-world HVAC systems, but relied on powerful desktop computers running costly numerical computation software such as MATLAB.

There are other studies using occupancy prediction for MPC in HVAC systems. For example, \cite{dong2009,dong2011-1,dong2011-2,dong2014} used machine learning to predict occupancy patterns based on environmental sensor data. More specifically \cite{dong2009} and \cite{dong2011-2} used predicted occupancy patterns to simulate HVAC control in EnergyPlus, whereas \cite{dong2011-1,dong2014} developed LTI MPC algorithms for HVAC control in real buildings. Furthermore, \cite{goyal2013} investigated the potential benefits of occupancy information for HVAC control, but their investigation was only limited to simulations.

The differences between our study and the aforementioned studies are: (1) we implemented our MPC algorithm in a real-world testbed using free-software platforms and low-cost embedded systems; (2) we employed EnergyPlus for real-time simulation-guided MPC of HVAC systems; (3) we developed a comprehensive solution that integrates occupancy recognition, occupancy prediction and simulation-guided MPC, thereby demonstrating the viability of automatic HVAC control using low-cost embedded systems.

\section{Occupancy Recognition} \label{sec:recognition}

\noindent Occupancy information is crucial for many applications, such as building management and human behavior studies.  We develop an occupancy recognition system based on real-time video processing of a video stream to infer the occupancy patterns dynamically. This raises a number of challenges. First, structures differ from one building to another, leading to the possibility of occupants being obscured by various obstacles, such as pillars. Hence, we need to track the movements of occupants to determine the occupancy more accurately.  Second, our algorithm is executed on an embedded system (e.g., Raspberry Pi), which has limited processing power and memory space compared to a typical desktop computer; this is particularly challenging since the typical video-processing algorithms are computationally demanding.

The basic idea of our occupancy-recognition algorithm is to count the number of people crossing a virtual reference line in the video, captured by a fisheye camera. Objects are identified as moving blobs (i.e., a set of connected points whose position is changing during the video stream); every such blob is interpreted as a person. Whenever a moving blob is detected, the algorithm keeps track of its movement to determine whether it passes the virtual reference line, and then updates the number of occupants accordingly. In particular, we position the line near the entrance of the space. Whenever the moving blob passes inward, the total number of occupants is increased by 1, and whenever the moving blob passes outward, the total number of occupants is decreased by 1.

\begin{figure}[ht!]
  \centering
  \scalebox{0.75}{
  \begin{tikzpicture}[node distance = 1.5cm, auto]

    \node (init) [init] {Initialize Algorithm};
    \node (isol) [process, right of=init, node distance=4.7cm, text width=4.0cm] {Isolate background by comparing current and previous frames};
    \node (extr) [process, below of=isol, node distance=1.65cm, text width=3.6cm] {Extract silhouettes in current frame};
    \node (add) [process, below of=extr, node distance=1.45cm, text width=3.7cm] {Add each silhouette to a tracking list};
    \node (if) [decision, below of=add, text width=5.2cm, aspect=4, node distance=1.7cm] {If a silhouette passes a reference line};
    \node (upda) [process, left of=if, text width=2cm, node distance=4.7cm] {Update occupancy};
    \node (rese) [process, below of=if, text width=4.0cm, node distance=1.75cm] {Reset occupancy if \\inconsistency detected};

    \draw [arrow] (init) -- (isol);
    \draw [arrow] (isol) -- (extr);
    \draw [arrow] (extr) -- (add);
    \draw [arrow] (add) -- (if);
    \draw [arrow] (if) -- node[auto] {yes} (upda);
    \draw [arrow] (if) -- node[auto] {no} (rese);
    \draw [arrow] (rese.east) -- +(35pt,0) |- (isol.east);
    \draw [arrow] (upda) |- (rese);
  \end{tikzpicture}}
  \caption{Flowchart of our occupancy-recognition algorithm.}
  \label{fig:flowchart}
\end{figure}
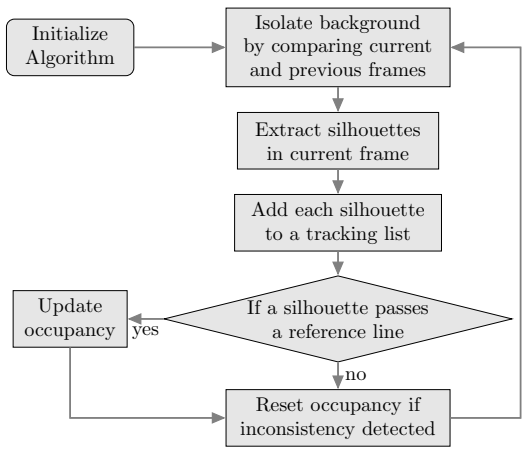

Our algorithm consists of the following five steps: (1) background isolation, (2) silhouette detection, (3) object tracking, (4) inward/outward logging, and (5) inconsistency resolution. The flowchart of our algorithm is presented in Figure~\ref{fig:flowchart}. In our implementation, two open-source projects were used: \emph{OpenCV} \cite{opencv_library} and \emph{openFramework} \cite{ox_library}. Next, we will explain each step in details.

\subsection{Background Isolation} \label{sub:background_updating}

\noindent The purpose of background isolation is to identify a background image in the current video frame. Note that the appearance of the background may vary over the course of the video, depending on the time-of-day (e.g., turning on the lights at night may significantly alter the appearance of the background compared to natural light). The shadows of occupants must also be taken into consideration during the background isolation. To overcome these challenges, we employ the Gaussian mixture-based background/foreground segmentation algorithm proposed in \cite{zivkovic2006efficient,zivkovic2004improved}, and implemented on OpenCV.

\subsection{Silhouette Detection} \label{sub:silhouette_searching}

\begin{figure}[ht!]
  \centering
  \includegraphics[width=.99\linewidth]{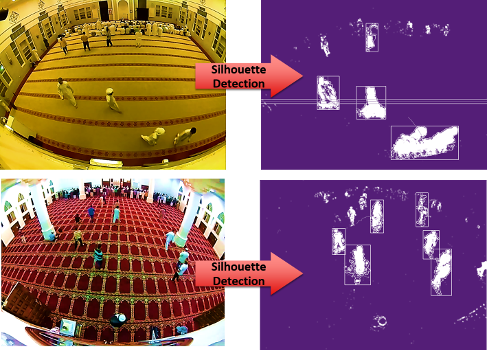}
  \caption{Examples of silhouette detection, taken from different buildings in which our system is deployed.}
  \label{fig:videoprocessing_result}
\end{figure}

\noindent In our setting, the term ``silhouette'' is used to refer to the border of a set of continuous points. Silhouette detection is based on the algorithm proposed in \cite{Suzuki85Topological}. The silhouettes of moving objects are extracted by comparing the current frame with the previous one; see Figure~\ref{fig:videoprocessing_result} for some examples. Here, only the silhouettes larger than a certain threshold area, $A$, are considered. The threshold is adjusted depending on the viewpoint and orientation of the camera. Furthermore, we use different values of $A$ for different parts of the space, to reflect the fact that individuals appear smaller as they move farther away from the camera. Importantly, the silhouettes of two or more people may overlap, and may, therefore, be interpreted as only one individual. To overcome this challenge, we use a machine-learning technique which will be explained later on in Section~\ref{sub:machine_learning}.

\subsection{Object Tracking}  \label{sub:object_tracking}

\noindent After obtaining the silhouettes of occupants, their bounding boxes are logged for tracking. To this end, the list ${T_a}$ of silhouettes from the last frame is maintained. The algorithm then obtains the list ${T_b}$ of silhouettes from the current frame. For each bounding box $B_b$ in ${T_b}$, the algorithm determines whether there exists a box $B_a$ in $T_a$ that is within a certain distance, $d$, from $B_b$. If so, then $B_b$ is assumed to be the new location of $B_a$. Consequently, $B_a$ is updated in $T_a$, and its location is set to be that of $B_b$, in preparation for the next iteration of the algorithm.  Note that the silhouettes are detected when they are moving. However, in cases where people might stop for a few seconds and then continue to walk, those objects will be removed from $T_a$ when they are missed for a certain number of seconds.

\subsection{Inward/Outward Logging}  \label{sub:in_out_checking}

\noindent Given the collection of moving objects and their locations, a virtual reference line is used to count occupants. Specifically, whenever the locations are updated, the algorithm checks every object to determine whether that object has crossed from one side of the line to the other. If so, then the number of occupants is updated according to the direction of the movement. For inward movement, the number of occupants increases by 1, otherwise it decreases by 1. Furthermore, since the silhouettes of any two moving objects may overlap, the width of the bounding box can be used to infer the number of occupants contained in each object.

\subsection{Inconsistency Resolution} \label{sub:error_detection}

\noindent With all the techniques described thus far, the performance of the algorithm may not be satisfactory, due to one major challenge: \emph{the silhouettes of different occupants may overlap}. In this case, some of the overlapping occupants may go undetected by the algorithm. This problem becomes even more evident when the movement patterns are affected by whether the occupants are entering or leaving the building, e.g., due to the fact that occupants arrive one by one, but leave all at once. For instance, in our application domain, the pace at which people leave is significantly faster than the pace at which they enter. Consequently, when all occupants leave at once, the number of overlapping silhouettes increases, leading to a larger number of people going undetected by the algorithm. As a result, the algorithm on average misses more occupants leaving than entering the building, leading to the erroneous conclusion that there are still occupants in the building when in fact there are none.

To resolve such inconsistency, two simple techniques are used. First, if the number of occupants becomes negative, the occupant counter is frozen until another object crosses the reference line inward. Second, if no moving object is detected for a certain period of time, the occupant counter is reset to zero. While these simple techniques reduce the error, they are clearly insufficient. In Section~\ref{sub:machine_learning}, we propose a dedicated technique to address this issue.

\subsection{Improvement by Machine Learning} \label{sub:machine_learning}

\noindent As mentioned earlier, one of the major challenges that we have encountered while deploying our occupancy-detection algorithm is to resolve the problem of overlapping silhouettes. One solution is based on the width of bounding box; the wider the box, the more occupants it contains. However, we observe that such a solution is insufficient, especially when an occupant happens to be directly in front of another. To resolve this issue, we employ machine-learning and image-classification techniques, coupled with randomized principal component analysis (PCA) \cite{halko2011finding}, which is implemented in \cite{scikit-learn}.

\begin{figure}[ht!]
\centering
\includegraphics[width=0.75\linewidth]{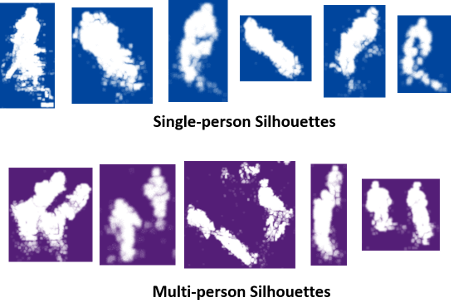}
\caption{Sample images from the training dataset used for machine learning and image classification.}
\label{fig:trainingdata}
\end{figure}

In more detail, we collected 13,000 blobs from video footages spanning a period of one week and segmented each such blob as a separate image to create our training dataset. The number of occupants in each image was manually identified; see Figure~\ref{fig:trainingdata} for some examples. The images were then transformed to grayscale in order to reduce the computational load. After that, the images were rescaled to the same size, 30$\times$15 pixels, thereby making the size of each image 450 pixels. Second, randomized PCA is used to project the pixels in the original array to a smaller array that preserves the characteristics of the images, as a set of 25 features for each image in this case. The projection aims to further reduce the computational complexity. Moreover, we added the original width, height and the ratio of black pixels to the set of features. After obtaining the features, Gaussian Naive Bayes is used to classify the blobs with respect to the number of occupants.

\subsection{Privacy Enhancement} \label{sub:privacy_concerns}

\begin{figure}[ht!]
\centering
\includegraphics[width=0.8\linewidth]{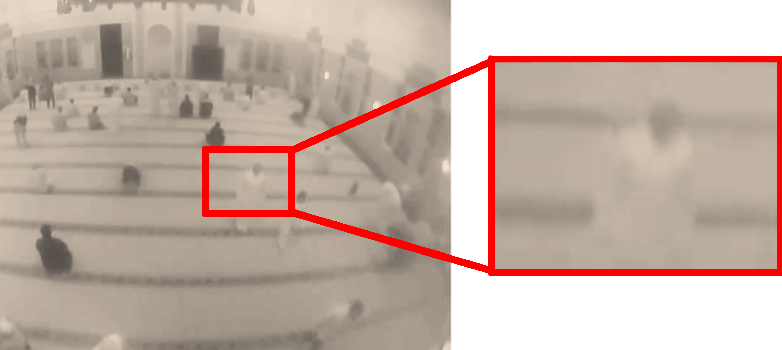}
\caption{Enhanced privacy by using frosted lens on camera.}
\label{fig:withoutlayer}
\end{figure}

\noindent In our system, the videos and images are discarded immediately after processing. Still, privacy invasion is always a concern for video based detection methods. With this in mind, we introduced a number of techniques to preserve the privacy of the occupants. First of all, to prevent any potential hacker from accessing the video feed, we ensured that the camera stream can only be accessed by a single process at a time. Consequently, when our system is running, all the other programs are blocked from accessing the camera.
In addition, a hardware-based solution has been tested. Specifically, we created a frosted lens by overlaying a semi-transparent layer in front of the camera, so as to blur the camera feed. In so doing, the faces of occupants are no longer recognizable; see Figure~\ref{fig:withoutlayer} for an illustration.

\subsection{Evaluation Results}  \label{sec:occupancy_detection_results}

\noindent In this section, we evaluate the accuracy of our occupancy recognition. To this end, we collected video footages from our testbed, each with a frame size of 800$\times$600 pixels, and a sample rate of 30 frames per second. The true occupancy was obtained by manually counting the number of occupants in each frame. Due to this laborious process, we were only able to obtain the true occupancy in a small number of videos, which are used in the evaluations.

We start by defining the accuracy rate as follows:
\begin{equation}
  {\sf AccuracyRate} \triangleq 1-\frac{{\sf Missing}_{\rm in}+{\sf Missing}_{\rm
  out}}{{\sf Total}_{\rm in}+{\sf Total}_{\rm out}}
\end{equation}
where ${\sf Total}_{\rm in}$ is the number of individuals who entered the building and ${\sf Missing}_{\rm in}$ is the number of such individuals who were undetected by the algorithm. Conversely, ${\sf Total}_{\rm out}$ is the number of individuals who exited the building and ${\sf Missing}_{\rm out}$ is the number of those who exited without being detected by our algorithm.

Table~\ref{tab:occupancyresults_1} presents the system accuracy given different numbers of occupants over different durations. As can be seen, the algorithm is able to recognize the number of occupants with high accuracy, even when over 400 individuals enter the building in just 20 minutes. Naturally, given a greater rate at which occupants enter the building, the accuracy of the system decreases because there are more cases in which the occupants overlap (in many such cases, it was hard, even for a human, to determine the exact number of occupants, and the video had to be replayed several times before the human was able to determine the true occupancy with certainty).

\begin{table}[ht!]
\centering
\caption{Results of occupancy recognition given different durations and different numbers of occupants.}
\label{tab:occupancyresults_1}
\scalebox{0.85}{
\begin{tabular}{@{}c|c|c@{}}
\hline\hline
Video Length     & Total No. of Occupants    & ${\sf AccuracyRate}$ \\
    & of Occupants &      \\ \hline
 40 mins  & 127          & 90\%             \\
 40 mins  & 154          & 90\%             \\
 20 mins  & 407          & 84\%             \\ \hline\hline
\end{tabular}}
\end{table}

Next, we evaluate our system over one-day periods during the weekend, weekday and Friday. Note that in our testbed Friday is the busiest day in the week due to the Friday sermon, which takes place only once a week, and typically attracts a large audience. The results are presented in Table~\ref{tab:occupancyresults_2}. As expected, the system accuracy correlates with the occupancy rates. In particular, the system is least accurate on Friday (which is the busiest day in our experiment), and most accurate during the weekend (which is the least busy in our experiment). Importantly, the accuracy seems sufficiently high throughout the week to provide a reasonable approximation of the actual occupancy state in the building, which is arguably sufficient for our purpose of HVAC control.

\begin{table}[ht!]
\centering
\caption{Results of occupancy recognition comparing weekend, weekday, and Friday.}
\label{tab:occupancyresults_2}
\scalebox{0.85}{
\begin{tabular}{@{}c|c|c@{}}
\hline\hline
Type    & Video Length & ${\sf AccuracyRate}$ \\ \hline
Weekend & 1 day        & 88\%   \\
Weekday & 1 day        & 86\%   \\
Friday  & 1 day        & 81\%   \\ \hline\hline
\end{tabular}}
\end{table}

We now turn our attention to quantifying the loss in accuracy that occurs when using our privacy-preserving frosted lens from Section~\ref{sub:privacy_concerns}. The results of this evaluation are presented in Table~\ref{tab:occupancyresults_3}. As can be seen, the frosted lens only reduces the accuracy slightly, and that is despite the fact that the video footage is considerably blurred, as we have shown earlier in Figure~\ref{fig:withoutlayer}.

\begin{table}[htbp!]
\centering
\caption{Results of occupancy recognition comparing normal lens and frosted lens.}
\label{tab:occupancyresults_3}
\scalebox{0.85}{
\begin{tabular}{@{}c|c|c|c@{}}
\hline\hline
Type          & Video Length  & Total No.     & ${\sf AccuracyRate}$ \\
              &     & of Occupants  &      \\ \hline
Normal Lens   & 160 mins  & 322           & 87.8\%           \\
Frosted Lens  & 160 mins  & 339           & 80.2\%           \\ \hline\hline
\end{tabular}}
\end{table}

Next, we evaluate the effectiveness of our machine-learning technique from Section~\ref{sub:machine_learning}. To this end, we use three performance measures that are widely used in the Machine-Learning literature. The first is ${\sf Precision}$, which is defined as the fraction of occupants that were correctly classified out of all those who were classified as either moving inward or as moving outward. The second measure is ${\sf Recall}$, defined as the fraction of occupants that were correctly classified out of all those who actually entered the building or exited it. {\color{black} Since it is often possible to have a naive classifier that has a high ${\sf Precision}$ but a low ${\sf Recall}$ or vice versa, a better metric called ${\sf F1\mbox{-}Score}$ has been utilized in \cite{f1score1,f1score2}, which is basically a harmonic mean of ${\sf Precision}$ and ${\sf Recall}$}. Based on those three measures, Table~\ref{tab:occupancyresults_4} shows the results before and after applying our machine-learning technique. The evaluation is carried out using 10-fold cross-validation of our dataset of 13,000 blobs. As can be seen, according to the most important measure, namely ${\sf F1\mbox{-}Score}$, our machine-learning technique from Section~\ref{sub:machine_learning} significantly improves the performance of our system.

\begin{figure*}[ht!]
  \centering
    \includegraphics[width=1\linewidth]{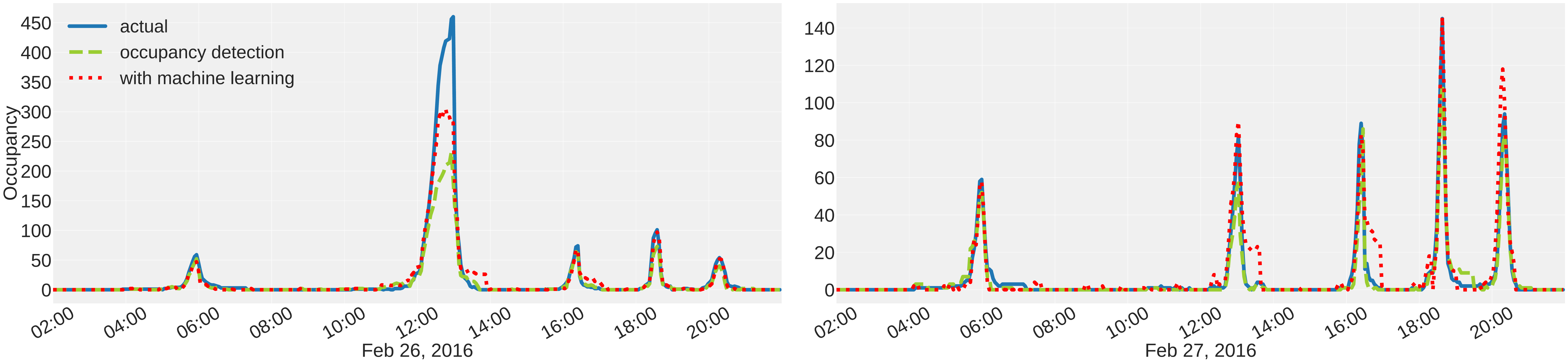}
    \caption{How occupancy changes during a typical day. Each peak represents one of the five daily prayers in Islam. We distinguish between Friday and other days of the week due to the Friday sermon, which precedes the midday prayer and usually attracts many more worshippers compared to any other prayer throughout the week (note that the scale of the y-axis is greater in the left plot than in the right plot).}\label{fig:occupancy_plot}
\end{figure*}

Finally, to better understand how occupancy changes during the daytime in our application, we plotted in Figure~\ref{fig:occupancy_plot} the actual occupancy, as well as the occupancy detected by our algorithm, given a typical Friday, and a typical Saturday. As can be seen, the general occupancy trend is clearly captured by the algorithm.

\begin{table}[ht!]
\centering
\caption{Results of occupancy recognition comparing with and without machine-learning technique.}
\label{tab:occupancyresults_4}
\scalebox{0.85}{
\begin{tabular}{@{}c|c|c|c@{}}
\hline\hline
Type                     & ${\sf Precision}$ & ${\sf Recall}$ & ${\sf F1\mbox{-}Score}$ \\ \hline
Without Machine Learning & 0.97              & 0.27           & 0.42             \\
With Machine Learning    & 0.69              & 0.78           & 0.73            \\ \hline\hline
\end{tabular}}
\end{table}

\section{Occupancy Prediction} \label{sec:prediction}
\noindent This section describes our methodology for predicting the future occupancy of the building. Such predictions can be very helpful when optimizing the HVAC control, especially in an application like ours, where occupants arrive in large numbers over short periods of time (see Figure~\ref{fig:occupancy_plot}). For example, being able to predict the arrival of a large number of people allows the system to pre-cool the building in anticipation of their arrival. Likewise, by predicting that all the occupants will shortly be leaving the building (e.g., if a certain social event was coming to an end), the system can turn off the HVAC system before the occupants even start departing.

In Section\ref{sub:the_general_approaches_for_predicting_the_occupancy_data} we propose a general-purpose approach to occupancy prediction. After that, in Section~\ref{sub:the_domain_specific_predicting_the_occupancy_data}, we further develop our occupancy prediction to produce a domain-specific approach, tailored to our application. Las\-tly, in Section~\ref{sub:how_to_evaluate_the_result_of_the_predicting_result} we evaluate our domain-specific prediction by quantifying the impact that it makes on the overall performance of our system.

\subsection{General Approach} \label{sub:the_general_approaches_for_predicting_the_occupancy_data}

\noindent As a starting point, let us consider a linear regression model, trained using all past occupancy data; let us denoted such an approach by $\widehat{\sf LR}_{\rm AllData}$. To take this simple approach one step further, let us extend the linear regression model by incorporating any ``\emph{special events}'' that may cause an abrupt change in the occupancy trend. Any such special event may be recurrent, with a slightly flexible timing and duration. For instance, consider a hall that can be booked in advance for social activities; here every such activity can be thought of as a \textit{special event}. Information about a special event (such as the average timing and duration, for example) may be available \emph{a priori} or may be inferred from the past occupancy data. The resulting approach, whereby special events are explicitly modeled, will be denoted by $\widehat{\sf LR}_{\rm SpEv}$, where ${\rm SpEv}$ stands for \emph{Special Event}. We suggest using this approach with the following linear regression model:

\begin{equation}\label{eq:lr_gen}
\widehat{y}^{(t)} = \beta_{\rm 0}+ \beta_{\rm 1} x^{(t)}_{\rm pastEvent} + \beta_{\rm 2} x^{(t)}_{\rm nextEvent} + \beta_{\rm 3} x^{(t)}_{\rm specialCase}
\end{equation}

\noindent where $\widehat{y}^{(t)}$ is the target variable (i.e., the predicted occupancy at time $t$); $\beta_{\rm i}:i\in\{0,1,2,3\}$ are the parameters of the model; $x^{(t)}_{\rm pastEvent}$ is the difference in time between $t$ and the timing of the \emph{past} event; and likewise $x^{(t)}_{\rm nextEvent}$ is the difference between $t$ and the timing of the \emph{next} event; $x^{(t)}_{\rm specialCase}$ is a binary variable that takes a value of 1 if $t$ coincides with a special case and takes a value of 0 otherwise (an example of such a binary variable is $x^{(t)}_{\rm newYearEve}$, which indicates whether $t$ coincides with new year's eve). Of course, Eq.~\ref{eq:lr_gen} is only meant as an example of the possible models that one could choose in order to explicitly model the special events that affect the occupancy. One may instead use other features, depending on the application at hand.

\subsection{Domain-specific Approach} \label{sub:the_domain_specific_predicting_the_occupancy_data}

\noindent In this section, we tailor our general-purpose approach from Section~\ref{sub:the_general_approaches_for_predicting_the_occupancy_data} to our testbed, i.e., the mosque. In this particular application, the \emph{special events} are the five daily prayers in Islam, the timing of which depends on the position of the sun in the sky. More specifically, the daily prayers are held (1) at dawn; (2) at midday right after the sun passes its highest; (3) at the late part of the afternoon; (4) just after sunset; and (5) between sunset and midnight. As such, the prayer times vary over the course of the year, because days tend to be longer in summer and shorter in winter. Importantly, the ``\emph{Friday prayer}'' (which takes place every Friday at midday) is preceded by a sermon, the attendance of which is obligatory for Muslims. As such, the number of occupants increases significantly compared to any other prayer throughout the week.

With this in mind, we now modify the general-purpose approach from Section~\ref{sub:the_general_approaches_for_predicting_the_occupancy_data} by incorporating our domain-specific knowledge of the application at hand. To this end, we consider (i) the difference in minutes between the current time and the \emph{past} prayer, (ii) the difference in minutes between the current time and the \emph{next} prayer, (iii) the current day of the week, and (iv) whether or not it is a public holiday. Then, to predict the occupancy at any given time, $\bar{t}$, we consider historical data for which the timestamp is within a certain threshold from $\bar{t}$. This threshold varies according to time of the year, to reflect the constantly-shifting prayer times. Based on these features, we propose the following linear regression model:

\begin{align}\label{eq:lr_context}
  \begin{split}
    \widehat{y}^{(t)} & = \beta_{\rm 0} + \beta_{\rm 1} x^{(t)}_{\rm pastPrayer} + \beta_{\rm 2} x^{(t)}_{\rm nextPrayer} \\ & + \beta_{\rm 3} x^{(t)}_{\rm holiday} + \beta_{\rm 4} x^{(t)}_{\rm dayOfWeek}
  \end{split}
\end{align}
where $x^{(t)}_{\rm pastPrayer}$ is the difference between $t$ and the timing of the \emph{past} prayer; $x^{(t)}_{\rm nextPrayer}$ is the difference between $t$ and the timing of the \emph{next} prayer; $x^{(t)}_{\rm holiday}$ is a binary variable indicating whether or not $t$ coincides with a public holiday; and $x^{(t)}_{\rm dayOfWeek}$ is the day of the week at time $t$. The resulting approach will be denoted by $\widehat{\sf LR}_{\rm DomSp}$, where ${\rm DomSp}$ stands for \emph{Domain Specific}. In addition to this linear regression model, we also experimented with a polynomial regression model, denoted by $\widehat{\sf PR}_{\rm DomSp}$, which uses the same features as those used in $\widehat{\sf LR}_{\rm DomSp}$.

\subsection{Evaluation Results} \label{sub:how_to_evaluate_the_result_of_the_predicting_result}

\noindent We use two standard metrics to evaluate the effectiveness of the proposed prediction models: ${\sf R\mbox{-}squared}$ and ${\sf RMSE}$ (Root Mean Squared Error). Both of these metrics are always between 0 and 1. Importantly, with ${\sf R\mbox{-}squared}$ the \emph{greater} the value the better the model, whereas with ${\sf RMSE}$ the \emph{smaller} the value the better the model.

\begin{figure}[ht!]
  \includegraphics[width=.99\linewidth]{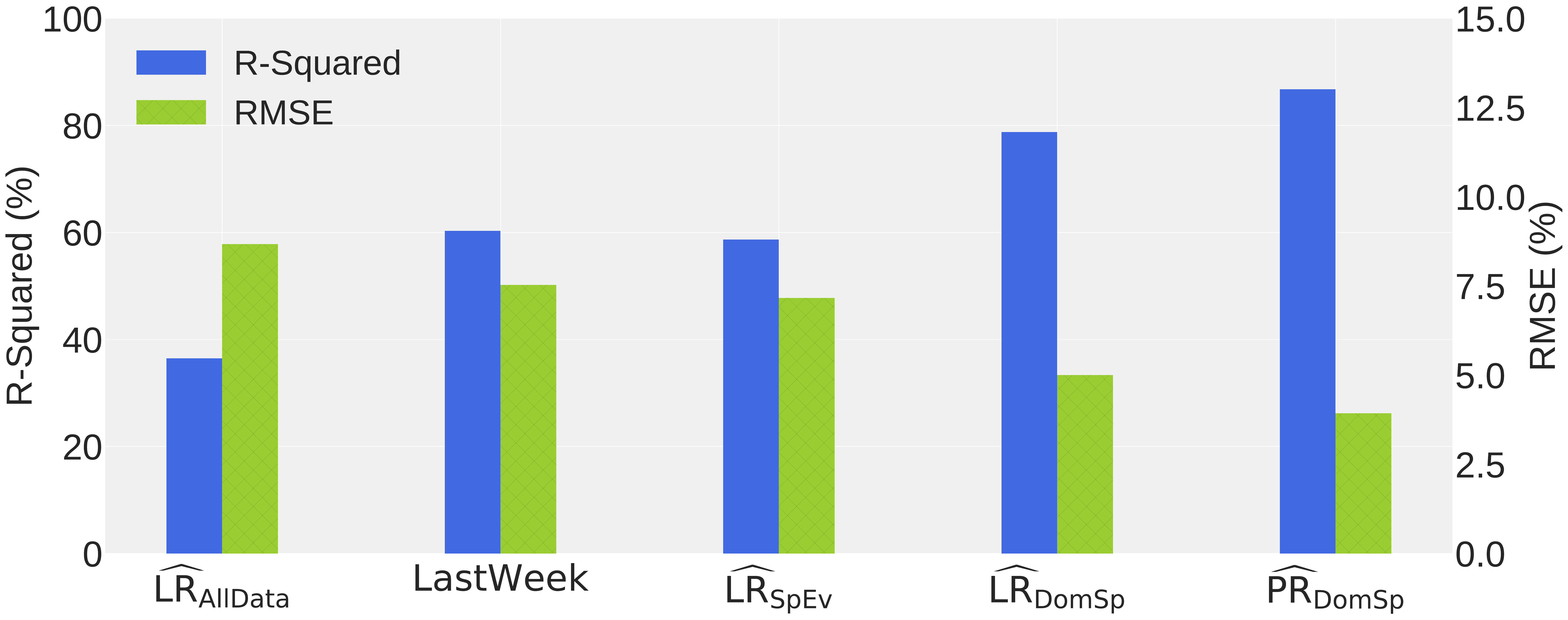}
  \caption{Results of occupancy prediction.}
  \label{fig:prediction_result}
\end{figure}

As a baseline model, we use a naive approach, denoted by {\sf LastWeek}, whereby the current week's occupancy is assumed to be identical to last week's occupancy (e.g., the occupancy pattern in, say, the coming Monday is assumed to be identical to that of last Monday). The results are shown in Figure~\ref{fig:prediction_result}, where all models are evaluated with a forecasting horizon of 24-hour ahead. As can be seen, even {\sf LastWeek} (the most naive approach) outperforms $\widehat{\sf LR}_{\rm AllData}$ (the general-purpose linear regression model trained using all past occupancy data).

In contrast, the performance of $\widehat{\sf LR}_{\rm SpEv}$ (the linear regression model whereby the special events are explicitly modeled) appears to be on par with that of {\sf LastWeek}. As for the domain-specific approaches, namely $\widehat{\sf LR}_{\rm DomSp}$ and $\widehat{\sf PR}_{\rm DomSp}$, they outperform the other alternatives due to three main reasons: (i) they take into account the day of the week, which is particularly important since the weekly sermon takes place every Friday; (ii) they are able to recognize public holidays, during which the occupancy typically increases in residential areas and decreases in commercial areas; (iii) they are trained using data from only the past 30 days (as opposed to using all past occupancy data), which is important since the prayer times are constantly shifting throughout the year; this shift is usually negligible over a 30-day period, but can be significant over the course of a year. Perhaps not surprisingly, between the two domain-specific approaches, $\widehat{\sf PR}_{\rm DomSp}$ outperforms $\widehat{\sf LR}_{\rm DomSp}$. Based on this evaluation, we adopt $\widehat{\sf PR}_{\rm DomSp}$ as our model of choice. With an R-squared value of about 0.87 and an RMSE value of about 0.03, this model appears to capture the overall occupancy trend to a satisfactory degree for the purpose of HVAC control.

\section{Building Thermal Response Simulation} \label{sec:simulation}

\noindent Traditional building modeling and control is often based on simplified mathematical building models due to their tractability and ease of use. However, these simplified models (e.g., first-principle linear models) can capture only limited aspects of the dynamic nature of buildings and systems. On the contrary, sophisticated building simulation software can use more realistic physical models, which in turn provide accurate modeling of the thermal behavior of buildings. One prominent example is \emph{EnergyPlus}, a popular building energy simulation program that can create detailed building envelope and system models. EnergyPlus can perform extensive conductivity analyses, and can even consider local weather conditions, by either incorporating user-supplied information or by using {\em EnergyPlus Weather Files} \cite{eplus_weather}. Furthermore, it provides interfaces with which external tools and programs can inter-operate. This process is known as \emph{co-simulation}, whereby different subsystems are integrated to carry out the simulation and calibration process simultaneously. With co-simulation, multiple simulators and software tools can be coupled in such a way that the strengths of each tool are exploited while overcoming their individual weaknesses. Currently, EnergyPlus is commonly used for planning and energy auditing. Using EnergyPlus for real-time HVAC control presents both opportunities and challenges. In this section, we utilize a framework of co-simulation in order to adapt EnergyPlus for real-world HVAC control.

\subsection{Basic Building Geometry}

\noindent First, the basic geometry of the building is created for EnergyPlus. In our testbed, we consider the large indoor space of a mosque installed with packaged rooftop HVAC units with ceiling-based cool air distribution. The detailed descriptions of the testbed, including dimensions and HVAC system specifications are provided in Table~\ref{tab:building-details}.

\begin{table}[htb!]
  \caption{Description of the testbed building.}
  \label{tab:building-details}
  \centering
  \scalebox{0.90}{
  \begin{tabular}{ c | c | c } \hline \hline
    Dimensions & \multicolumn{2}{c}{18m$\times$18m$\times$7m} \\  \hline
    \multirow{3}{*}{Walls} & Outer Layer	& Stucco \\
                           & Middle Layer & Concrete \\
                           & Inner Layer  & Gypsum \\ \hline
    Doors & \multicolumn{2}{c}{Wood Stile and Rail} \\ \hline
    Windows & \multicolumn{2}{c}{Glass with Vinyl Framing} \\ \hline
    \multirow{2}{*}{Packaged Rooftop HVAC}  & Cooling Capacity & 175840W \\ 
          & Air Flow Rate & 7.56m$^3$/sec \\ \hline \hline
   \end{tabular}}
\end{table}

\begin{figure}[ht!]
\centering
\begin{subfigure}[t]{0.5\linewidth}
  \centering
  \includegraphics[width=0.99\linewidth]{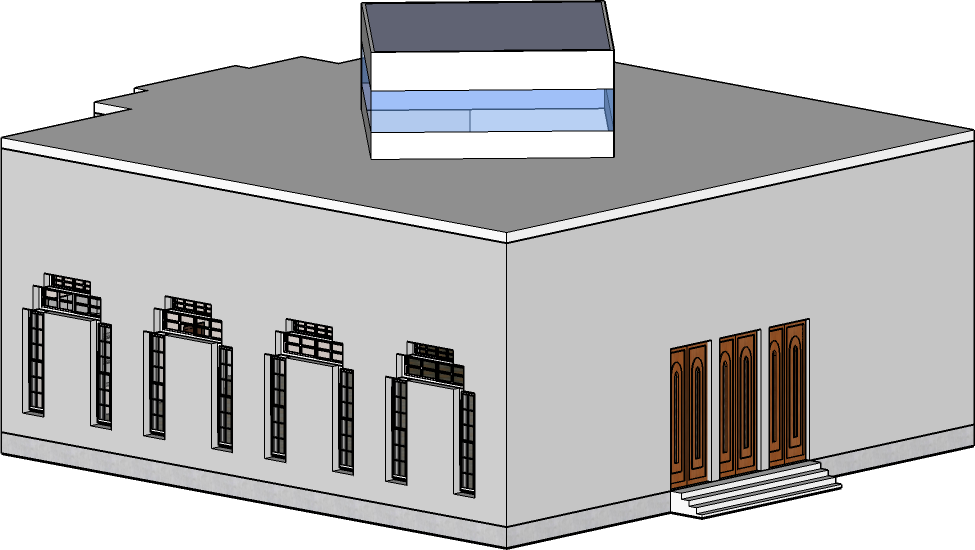}
  \caption{}
  \label{fig:sim_sketchup}
\end{subfigure}%
\begin{subfigure}[t]{0.5\linewidth}
  \centering
  \includegraphics[width=0.99\linewidth]{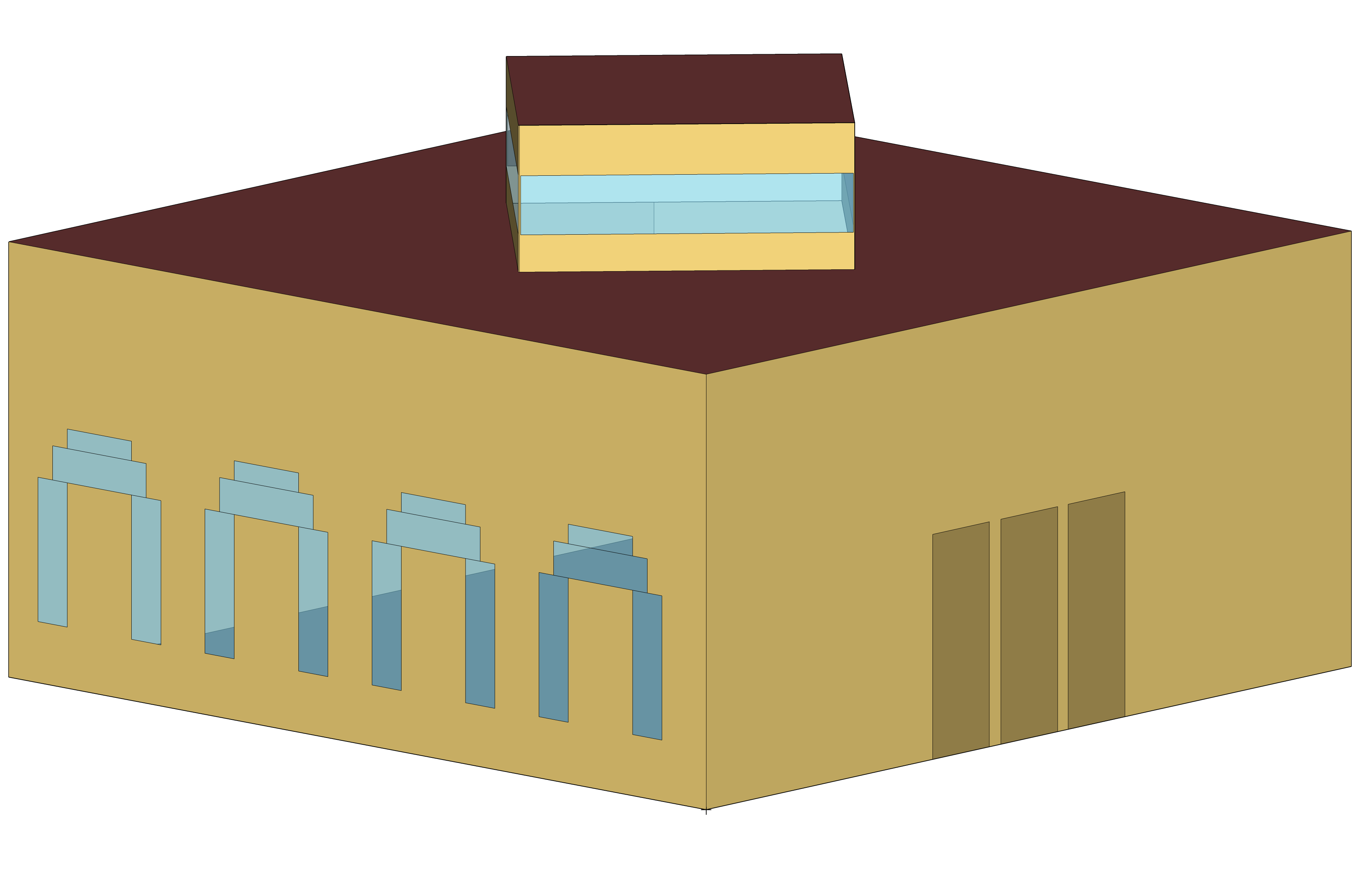}
  \caption{}
  \label{fig:sim_hvac}
\end{subfigure}
\caption{Basic building models. (a) 3D \emph{SketchUp} building model, (b) the corresponding \emph{OpenStudio} model.}
\label{fig:sim_model}
\end{figure}

We employ \emph{SketchUp} for 3D modeling with the default material properties for walls, windows, and doors. Then, the 3D model is imported to \emph{OpenStudio}, where the parameters of the HVAC system are incorporated into the model, based on the vendor's specifications and datasheet. The SketchUp building model and the corresponding OpenStudio model are visualized in Figure~\ref{fig:sim_model}. Apart from the basic geometry and properties of the building, many parameters still need to be calibrated according to the available data; this will be the focus on the next subsection.

\subsection{Building Model Calibration}\label{sec:buildingModelCalibration}

\begin{figure}[ht!]
  \centering
  \includegraphics[width=0.95\linewidth]{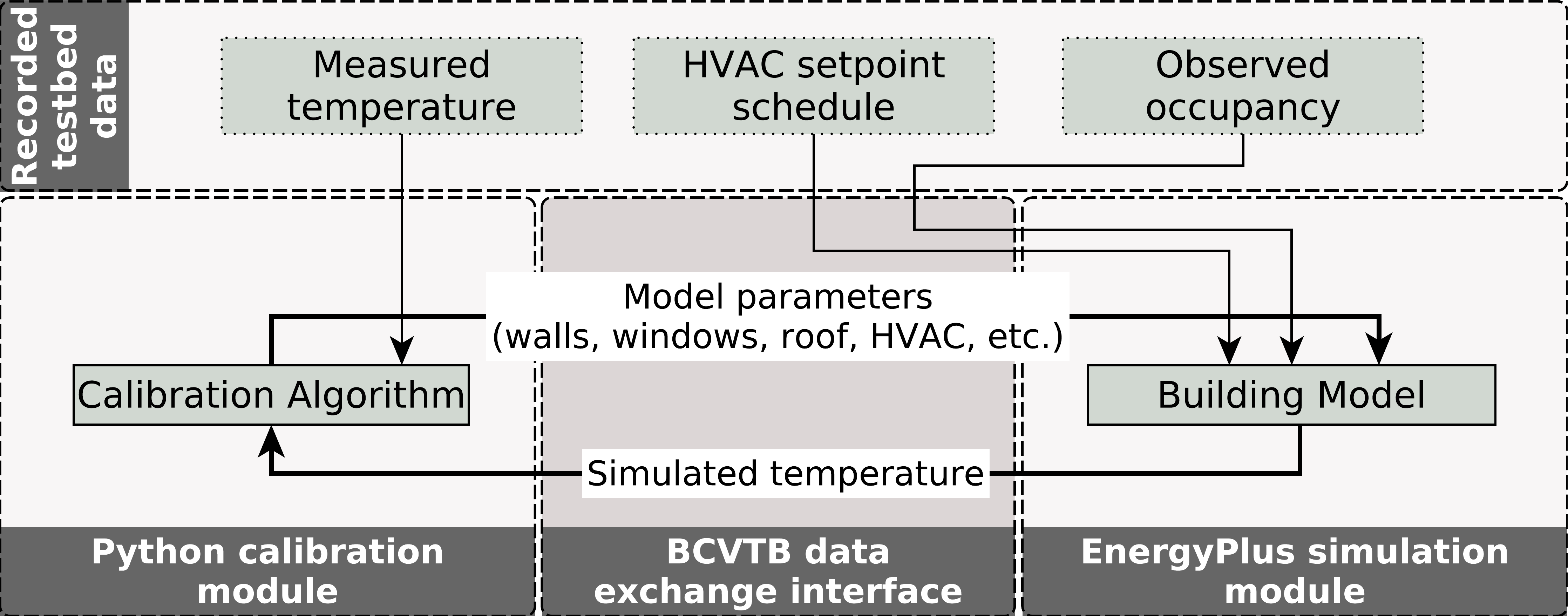}
  \caption[]{Co-simulation framework using BCVTB.}
  \label{fig:sim_framework}
\end{figure}

\noindent To obtain an accurate simulation of the thermal response of the building, we must first calibrate the parameters of the building model so as to minimize the gap between the simulated indoor temperature and the measured one. To this end, we use an inverse calibration approach, whereby the model's outputs are used to calibrate the parameters of the model. This calibration process is carried out using the co-simulation framework illustrated in Figure~\ref{fig:sim_framework}. In particular, the framework couples the EnergyPlus building model with our calibration algorithm which we implemented using \emph{Python}; this coupling is done using the BCVTB (Building Controls Virtual Test Bed) software tool \cite{bcvtb}, which provides a data-exchange interface between EnergyPlus and Python. Moreover, BCVTB allows EnergyPlus to take into consideration the real-world data measured from our testbed, such as observed occupancy and temperature.
 
Out of the numerous parameters that can be calibrated in EnergyPlus, we focus on a particular set of uncertain parameters as our candidates for calibration, because they are either unknown or likely to deviate from the datasheet. Specifically, these parameters are the cooling capacity and air flow rate of the HVAC system, as well as the thickness and conductivity of the roof, the walls, and the windows.

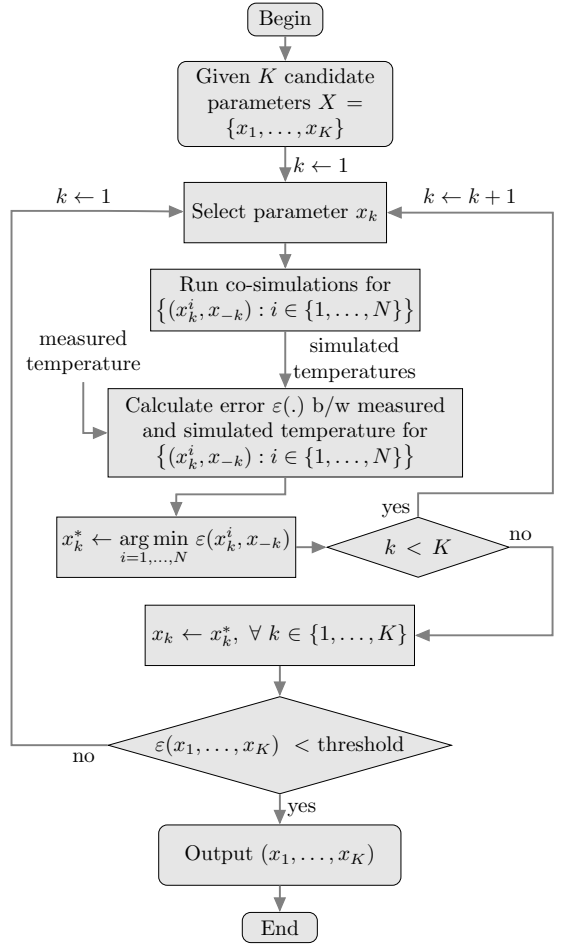
\begin{figure}[ht!]
  \centering
  \scalebox{0.8}{
  \begin{tikzpicture}[node distance = 1.5cm, auto]

    \node (start) [startstop] {Begin};
    \node (init) [init, below of=start, text width=3.3cm, node distance=1.4cm]     {Given $K$ candidate \\ parameters $X=\{x_1,\ldots,x_K\}$};
    \node (select) [process, below of=init, text width=3.4cm, node distance=1.8cm]    {Select parameter ${x_k}$};
    \node (run) [process, below of=select, text width=4.5cm, node distance=1.45cm]     {Run co-simulations for $\left\{(x_k^i, x_{-k}) : i\in\{1,\ldots,N\} \right\}$};
    \node (error) [process, below of=run, text width=5.9cm, node distance=2.2cm]    {Calculate error $\varepsilon(.)$ b/w measured \\ and simulated temperature for \\ $\left\{(x_k^i, x_{-k}) : i\in\{1,\ldots,N\}\right\}$};
    \node (minimum) [process, below left= 0.65cm and -3.1cm of error, text width=4.0cm]     {$x^*_k \gets \underset{i=1,\dots,N}{\arg\min} \ \varepsilon (x_k^i, x_{-k})$};
    \node (input) [input, above left = 0.10cm and +0.40cm of error] {};
    \node (if1) [decision, right=0.5cm of minimum, aspect=3.0, text width=2.0cm] {$k<K$};
    \node (assign) [process, below left=0.7cm and -0.7cm of if1, text width=4.5cm] {$x_k\gets x^*_k, \ \forall \ k\in\{1,\ldots,K\}$};
    \node (if2) [decision, below=0.5cm of assign, aspect=3.5, text width=6.0cm]    {$\varepsilon(x_1, \dots, x_K)<$ threshold};
    \node (done) [init, below= 0.5cm of if2, text width=3.75cm]     {Output $(x_1,\dots,x_K)$};
    \node (end) [startstop, below of=done, node distance=1.2cm]     {End};

    \draw [arrow] (start) -- (init);
    \draw [arrow] (init) -- node[align=center] {$k \leftarrow 1$} (select);
    \draw [arrow] (select) -- (run);
    \draw [arrow] (run) -- node[align=center] {simulated\\temperatures} (error);
    \draw [arrow] (input) -- +(0,  0) node[align=center,near start,above] {measured\\temperature} -- +(0, -24pt) -- (error.west);
    \draw [arrow] (error.south) -- +(0,-8pt) -- +(-51pt,-8pt) -- (minimum.north);
    \draw [arrow] (minimum) -- (if1);
    \draw [arrow] (if1.east) -- +(20pt,0) node[near start] {no} |- (assign);
    \draw [arrow] (assign) -- (if2);
    \draw [arrow] (if1.north) -- +(0, 11pt) node[near start] {yes} -- +(63pt, 11pt) |- node[above, near end] {$k \leftarrow k+1$}  (select);
    \draw [arrow] (if2.west) -- +(-45pt, 0) node[near start] {no} |- +(-45pt, 252pt) -- +(0, 252pt) node[near end] {$k \leftarrow 1$} -- (select);
    \draw [arrow] (if2) -- node[align=center] {yes} (done);
    \draw [arrow] (done) -- (end);
  \end{tikzpicture}}
  \caption{Flowchart of the model-calibration algorithm.}
  \label{fig:sim_calibration_flowchart}
\end{figure}

Now, we will explain how the calibration process is done. To this end, we use a dedicated algorithm, the flowchart of which is depicted in Figure~\ref{fig:sim_calibration_flowchart}. The algorithm is based on the notion of gradient descent, where values of those parameters are updated iteratively until the error between the simulated indoor temperature and the measured temperature is within an acceptable threshold. Next, we explain each step of this algorithm.

Let $X=\{x_1,x_2,\dots,x_K\}$ denote the set of parameters to be calibrated. For every $x_k\in X$, the algorithm runs a series of $N$ co-simulations, updating the model in every iteration using a different value of $x_k$ while keeping the values of the remaining parameters unchanged. Specifically, in the $i^{\textnormal{th}}$ iteration of this process (where $i\in\{1,\ldots,N\}$) the error is calculated as the difference between the measured temperature and the simulated model temperature; this error is denoted by:
$$
  \varepsilon(x_1,\ldots, x_{k-1},x_k^i,x_{k+1},\ldots,x_K)
$$

\noindent where $x_k^i$ takes the value of $x_{k}$ in the $i^{\textnormal{th}}$ iteration. For notational convenience, let us write $(x^i_k, x_{-k})$ instead of writing $(x_1,\ldots, x_{k-1},x^i_k,x_{k+1},\ldots,x_K)$. Then the optimal value for $x_k$, denoted by $x^*_k$, can be computed as follows:
\begin{equation}\label{eq:calib2}
  x^*_k = \underset{i=1,\dots,N}{\arg\min} \ \ \varepsilon (x^i_k, x_{-k})
\end{equation}

After computing $x^*_k$ for every $k\in\{1,\ldots,K\}$, the algorithm checks whether $\varepsilon(x^*_1, x^*_2, \dots, x^*_K)$ is within the acceptable threshold. If so, then it outputs $(x^*_1, x^*_2, \dots, x^*_K)$ and terminates; otherwise it repeats the entire process but after updating the parameters, i.e., after setting $x_k\gets x^*_k$ for every $k\in\{1,\ldots,K\}$. We note that the outcome of the calibration process is not affected by the order in which the algorithm iterates over the parameters.

\subsection{Evaluating the Model-Calibration Algorithm}
To evaluate the algorithm from Figure~\ref{fig:sim_calibration_flowchart}, we use two standard metrics, namely: the Coefficient of Variation of Root Mean Squared Error (${\sf CVRMSE}$) and the Mean Bias Error (${\sf MBE}$); these are computed as follows:

\begin{equation}
  {\sf CVRMSE} = \frac{\sqrt{\sum_{t=1}^{M}
  \left[\frac{(T_{\rm t}-\hat{T}_{\rm t})^2}{M}\right]}}{\frac{1}{M}\sum_{t=1}^{M}T_{\rm t}}
  \label{eq:rmse}
\end{equation}
\begin{equation}
  {\sf MBE} = \frac{\sum_{t=1}^{M}(T_{\rm t}-\hat{T}_{\rm t})}{\sum_{t=1}^{M}T_{\rm t}}
  \label{eq:mbe}
\end{equation}

\noindent where ${T_{\rm t}}$ and ${\hat{T}_{\rm t}}$ are the measured and simulated temperature values at time $t$, respectively, and $M$ is the total number of simulation time steps. Both {\sf MBE} and {\sf CVRMSE} provide different insights to the calibration process. However, the {\sf CVRMSE} is arguably superior because unlike {\sf MBE} it does not suffer from the cancellation effect \cite{cvrmse_mbe}.

\begin{table}[htb!]
  \small
  \caption{Parameters used for model calibration.}
  \label{tab:sim_params}
  \centering
  \resizebox{0.47\textwidth}{!}{%
  \begin{tabular}{@{} x{.095\textwidth} @{} |  @{} x{.110\textwidth} @{} |  @{} x{.110\textwidth} @{} |  @{} x{.120\textwidth} @{} |  @{} x{.120\textwidth} @{}} \hline \hline
    Parameter & \multicolumn{2}{c|}{Field} & Initial Value & Calibrated Value \\  \hline

    & Outer Layer	& Thickness & 2.53cm & 7.72cm \\ \cline{3-5}
    & (Stucco) & Conductivity & 0.69W/(m-K) & 0.16W/(m-K)\\ \cline{2-5}
    Exterior & Layer 2 & Thickness & 20.32cm & 62.01cm \\ \cline{3-5}
    Walls & (Concrete) & Conductivity & 1.31W/(m-K) & 0.311W/(m-K) \\ \cline{2-5}
    & Layer 3	& Thickness & 1.27cm & 3.87cm \\ \cline{3-5}
    & (Gypsum) & Conductivity & 0.16W/(m-K) & 0.037W/(m-K) \\ \hline
    \multirow{2}{*}{HVAC} & \multicolumn{2}{l|}{Cooling Capacity}	& 87920W & 164850W \\ \cline{2-5} & \multicolumn{2}{l|}{Air Flow Rate} & 3.78m$^3$/sec & 7.08m$^3$/sec \\ \hline \hline
   \end{tabular}}
\end{table}

\begin{figure*}[htb!]
  \centering
  \begin{subfigure}[b]{0.475\textwidth}
      \centering
      \includegraphics[width=\textwidth]{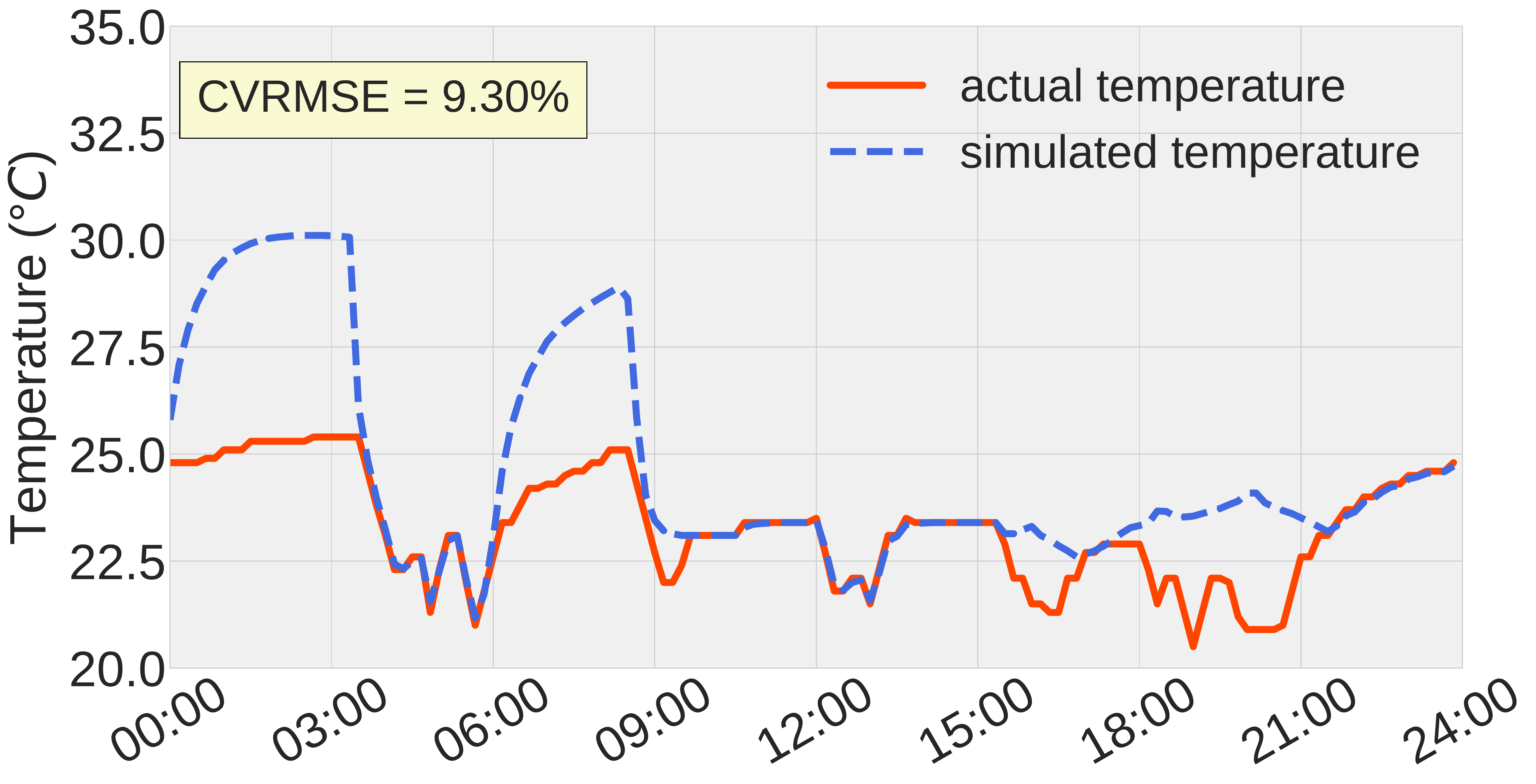}
      \caption{Using the \emph{default} parameter values set by \emph{SketchUp} during initial model creation.}
      \label{fig:sim_calibration1}
  \end{subfigure}
  \begin{subfigure}[b]{0.475\textwidth}
      \centering
      \includegraphics[width=\textwidth]{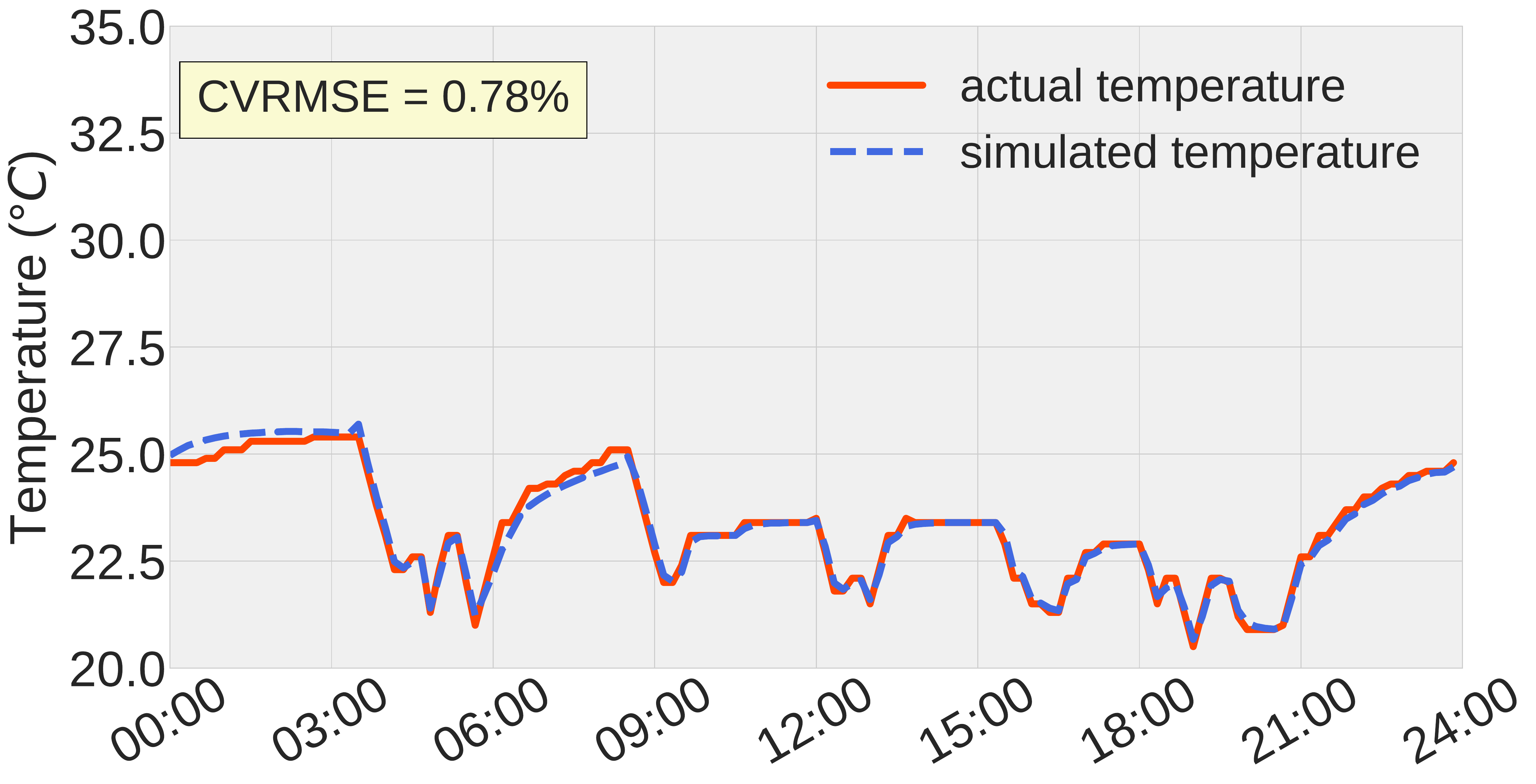}
      \caption{Using the \emph{calibrated} parameter values from our model-calibration algorithm.}
      \label{fig:sim_calibration2}
  \end{subfigure}
  \caption{Results of model calibration.}
  \label{fig:sim_calibration}
\end{figure*}

\begin{figure*}[htb!]
  \centering
  \begin{subfigure}[b]{0.475\textwidth}
      \centering
      \includegraphics[width=\textwidth]{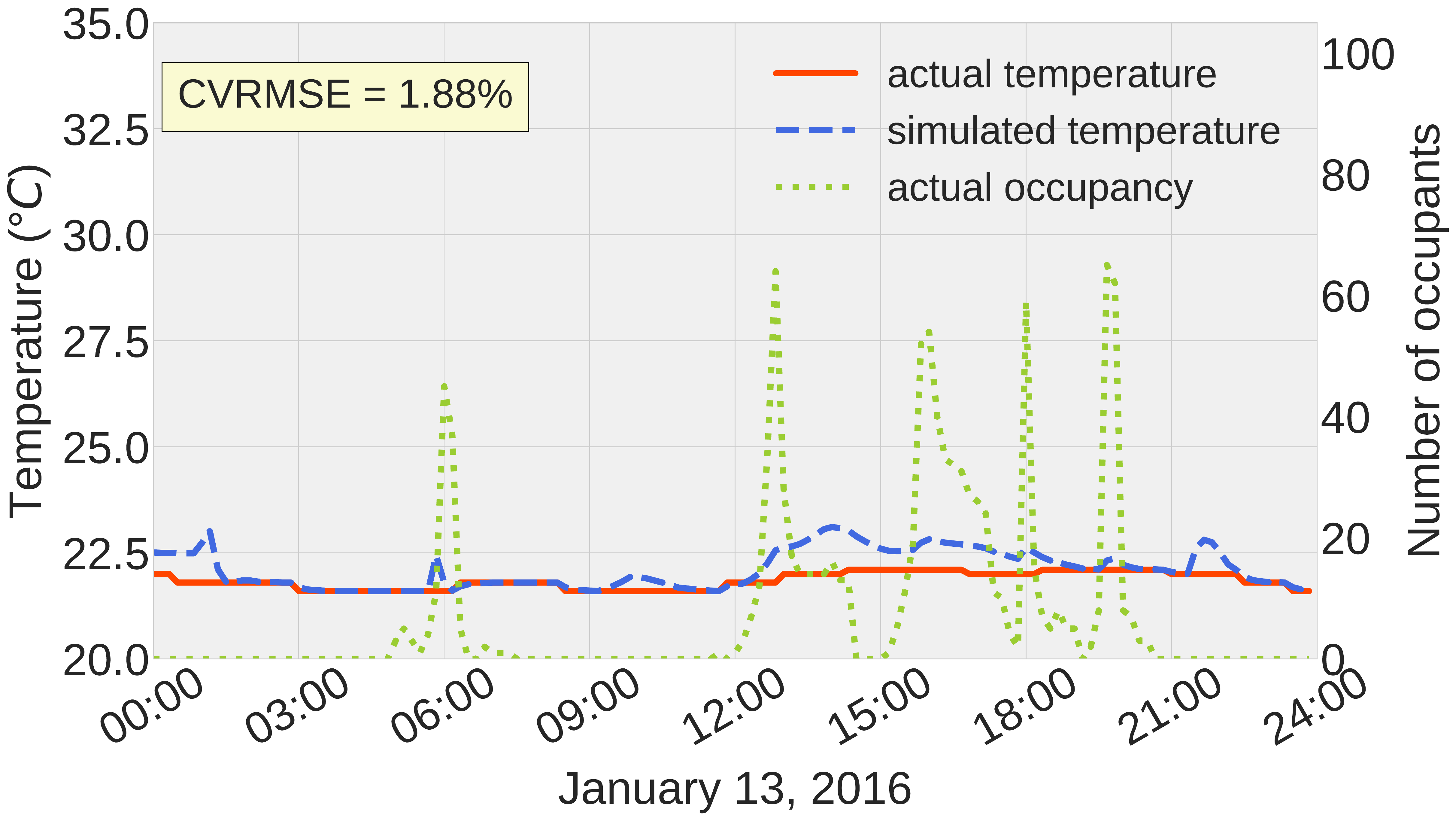}
      \caption{A typical day in winter with AC turned off.}
      \label{fig:sim_validation1}
  \end{subfigure}
  \begin{subfigure}[b]{0.475\textwidth}
      \centering
      \includegraphics[width=\textwidth]{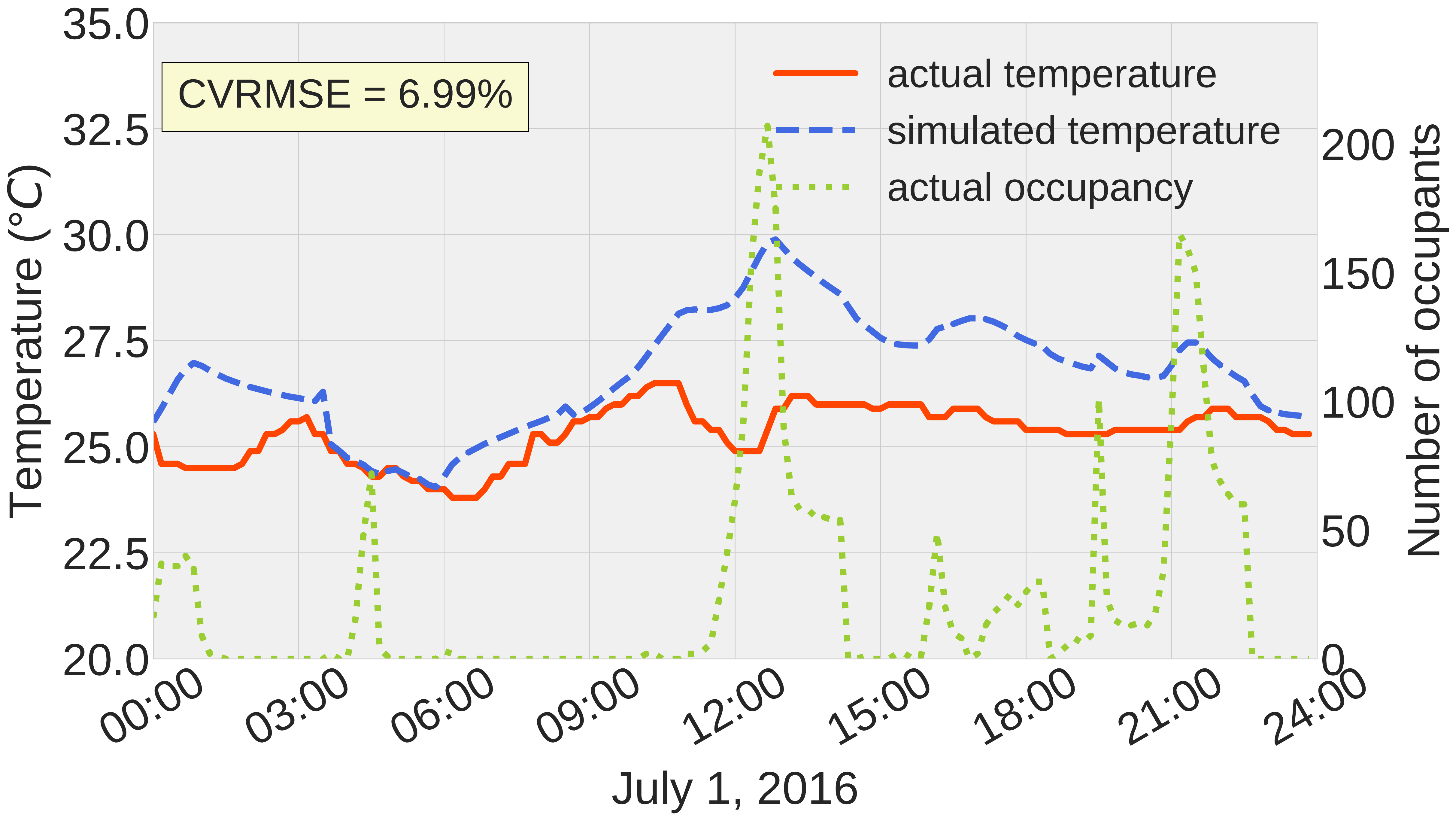}
      \caption{A hot day in summer.}
      \label{fig:sim_validation2}
  \end{subfigure}
  \caption{Results of testing calibrated model with validation sets.}
  \label{fig:sim_validation}
\end{figure*}

The default and calibrated values of our model parameters are listed in Table~\ref{tab:sim_params}. Note that we don not manually specify the allowed parameter ranges. EnergyPlus performs automatic parameter range checking using Input Data Dictionary \cite{idd}, which specifies the maximum and/or minimum values of parameters.

Using a simulation period of 24 hours, we tested the temperature response of our model before and after the calibration process (we used 24 hours since it is adequate for our purpose of HVAC control). Specifically, the calibration process takes around 100 co-simulations on average, each of which uses sub-hourly data with a sampling time period of 10 minutes. A CVRMSE of less than 2.0\% was treated as an acceptable calibration threshold. The results are depicted in Figure~\ref{fig:sim_calibration}. In particular, Figure~\ref{fig:sim_calibration1} plots the actual temperature as well as the simulated temperature, given the \emph{default} parameter values from \emph{SketchUp}. In Figure~\ref{fig:sim_calibration2}, the default parameter values are replaced by the ones calibrated using our algorithm from Section~\ref{sec:buildingModelCalibration}. As can be seen, the simulated temperature of the calibrated EnergyPlus model closely matches the measured temperature. Quantifying the performance gains using the aforementioned metrics, we find that the algorithm achieves {\sf MBE}$<$1\% and {\sf CVRMSE}$<$1\%. This seems satisfactory, bearing in mind that the ASHRAE guidelines, which require that {\sf MBE}$<$10\% and {\sf CVRMSE}$<$30\% for hourly data, and {\sf MBE}$<$5\% and {\sf CVRMSE}$<$15\% for monthly data \cite{ashrae14}. We tested our calibrated model with 24-hour periods taken from different times of the year to cover hot and cold days. The results are shown in Figure~\ref{fig:sim_validation}. In particular, Figure~\ref{fig:sim_validation1} shows the temperature response of the model on a typical winter day when the HVAC system is turned off. It can be seen that CVRMSE is still below the 2.0\% threshold and the model is considered calibrated. Figure~\ref{fig:sim_validation2} shows the model response on a hot summer day. In this case, CVRMSE has become above the threshold and the algorithm will need to re-calibrate the model to once again make CVRMSE below the threshold. These figures also show the detected occupancy to demonstrate how the model response varies with changing occupancy patterns. We note that in our testbed building there are no days without occupancy because there are always multiple prayers and multiple worshipers every day.

\begin{figure}[ht]
\centering
\begin{subfigure}[t]{0.5\linewidth}
  \centering
  \includegraphics[width=0.99\linewidth]{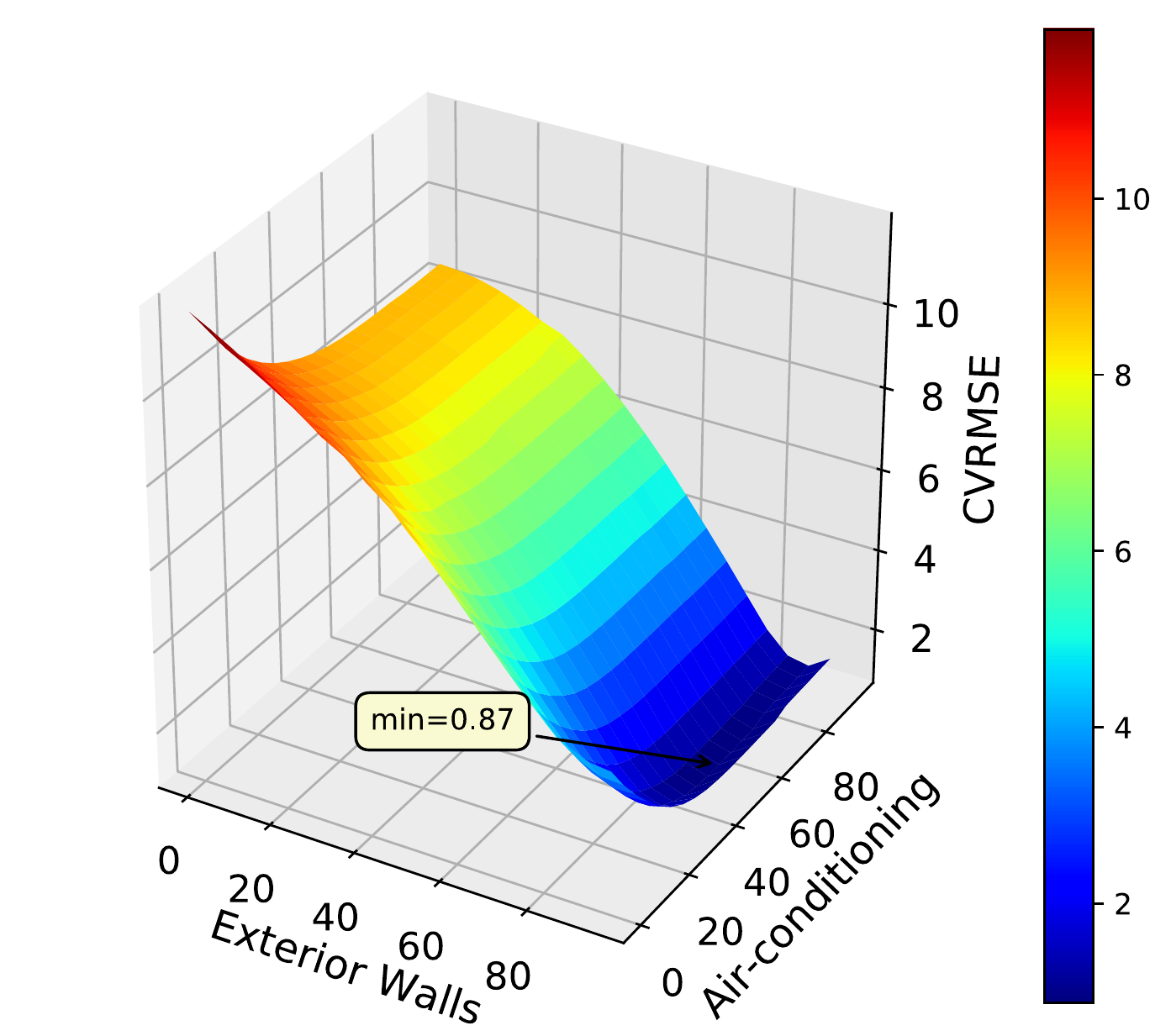}
  \caption{}
  \label{fig:sim_calibration_contour_a}
\end{subfigure}%
\begin{subfigure}[t]{0.5\linewidth}
  \centering
  \includegraphics[width=0.99\linewidth]{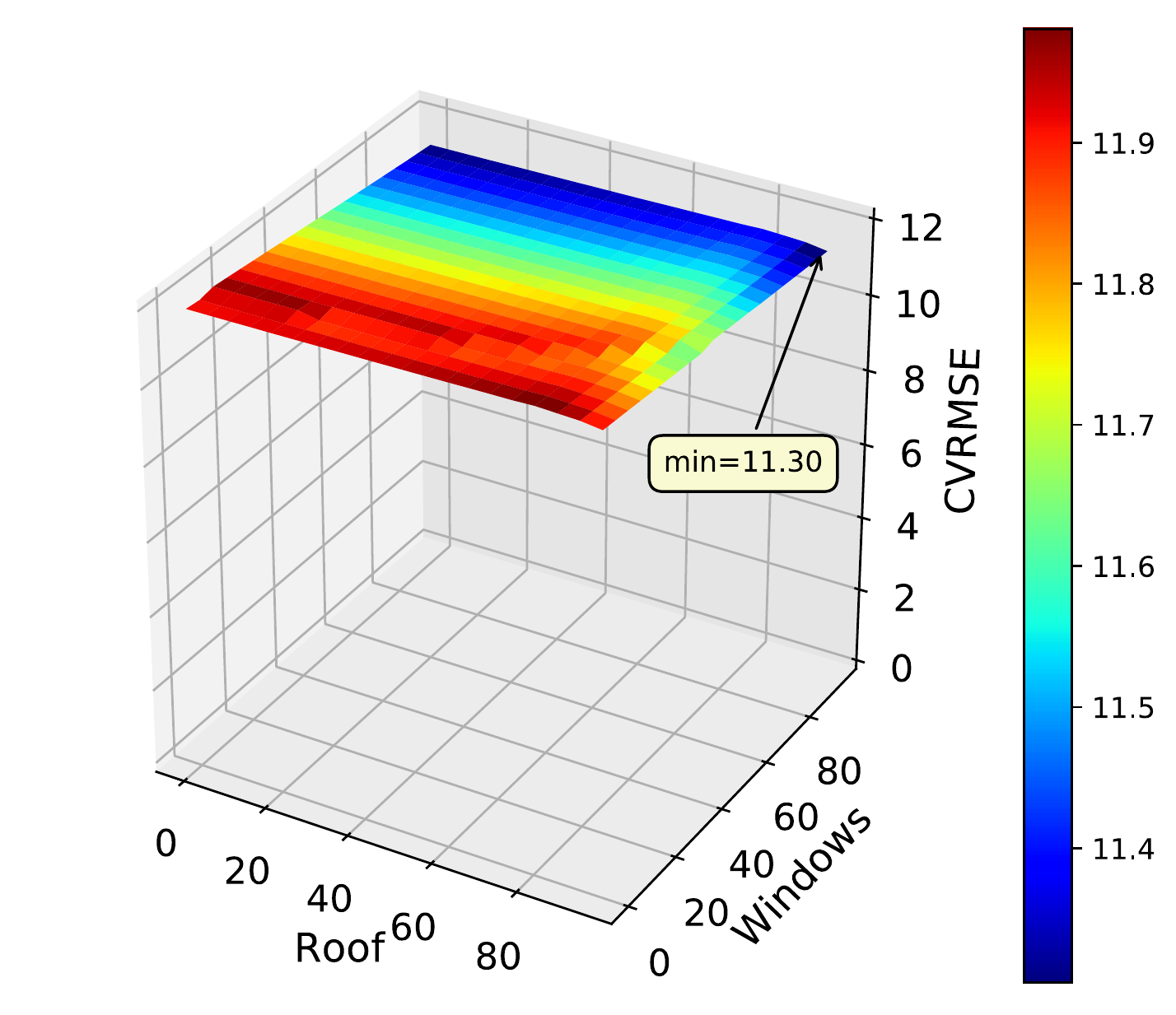}
  \caption{}
  \label{fig:sim_calibration_contour_b}
\end{subfigure}
\caption[]{Examples of how the performance improvements from the calibration process can vary significantly depending on the parameters being calibrated.}
\label{fig:sim_calibration_contour}
\end{figure}

We observe that the calibrated values of the uncertain HVAC parameters are very close to the manufacturer's specifications. In other words, even if those specifications were not available, our approach can still be applied without any noticeable change in performance. This makes our approach more practical in situations where such information is not available. We experimented with parameters other than those outlined in Table~\ref{tab:sim_params}, and found that their calibration did not yield any noticeable improvements. An example is illustrated in Figure~\ref{fig:sim_calibration_contour}. In particular, Figure~\ref{fig:sim_calibration_contour_a} depicts the impact of calibrating two parameters from Table~\ref{tab:sim_params}, whereas Figure~\ref{fig:sim_calibration_contour_b} depicts the impact of calibrating two parameters that are not outlined in Table~\ref{tab:sim_params}, which are: (i) the roof’s thickness and conductivity, and (ii) the windows’ thickness and conductivity. Here, the x-axis and y-axis represent the relative change (compared to the default value) in the first parameter and in the second parameter, respectively, while the z-axis represents the corresponding {\sf CVRMSE} value. As can be seen, the performance improvements from the calibration process vary significantly depending on the parameters being calibrated.

Finally, commenting on the runtime of the calibration algorithm, our implementation on Raspberry Pi 3 took an average of 43 seconds per co-simulation, which allows for about 2000 co-simulations per day (recall that our satisfactory results from Figure~\ref{fig:sim_calibration} required only 100 co-simulations). These results demonstrate that our calibration methodology can be used in practice, to automatically and continuously correct the building model for effective real-time HVAC control using low-cost embedded computers such as Raspberry Pi.

\section{Simulation-guided Model Predictive Control} \label{sec:eval}

\noindent Building upon our occupancy-prediction from Section~\ref{sec:prediction} and our temperature-response simulation from Section~\ref{sec:simulation}, we now propose a model predictive control algorithm in Section~\ref{subsec:control_framework}, and evaluate it in a real-world HVAC system in Section~\ref{subsec:eval_res}.

\begin{figure}[ht!]
  \centering
  \includegraphics[width=\linewidth]{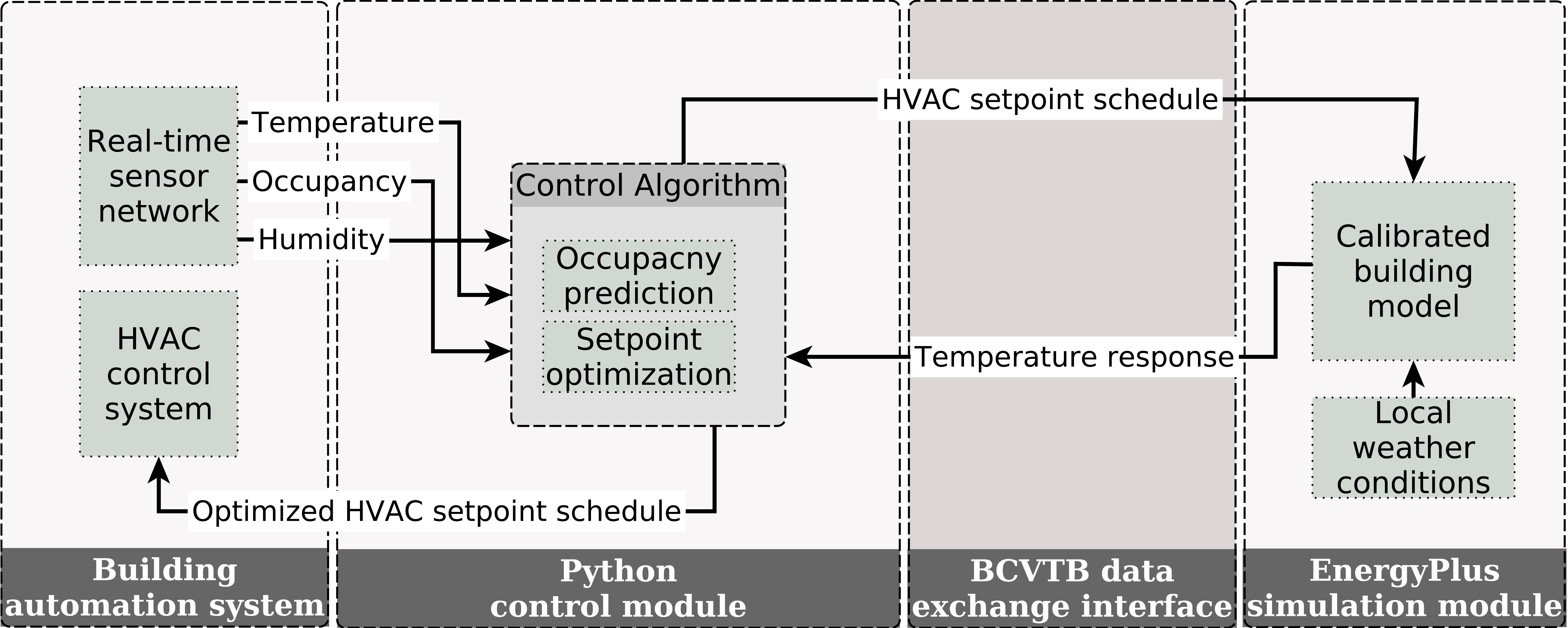}
  \caption{Framework of model predictive control algorithm.}
  \label{fig:eval_framework}
\end{figure}

\subsection{HVAC Control Framework} \label{subsec:control_framework}

\noindent The framework of our model predictive control consists of several parts that interact with one another as illustrated in Figure~\ref{fig:eval_framework}. {\color{black} Here, a control algorithm provides an HVAC setpoint schedule to the EnergyPlus simulator through the BCVTB interface. Based on this schedule and the local weather information, the EnergyPlus simulator performs simulations and provides temperature response to the control algorithm. Note that the local weather information is provided by the {\em EnergyPlus Weather File} \cite{eplus_weather} which includes temperature, humidity, pressure, wind speed, solar radiation, luminance, precipitation etc.} Now, based on the temperature response as well as the occupancy prediction, the control algorithm checks whether thermal-comfort criteria will be satisfied; if so, then there is no need for any modifications to the HVAC setpoint schedule; if not, then the HVAC setpoint schedule is modified subject to the thermal-comfort threshold, before being sent again to the EnergyPlus simulator, and so on. {\color{black} Once the HVAC setpoint schedule is finalized, the HVAC system starts following this schedule, and the control algorithm starts monitoring the real-time temperature, humidity, and occupancy data, to determine the actual thermal comfort that resulted from its control decisions, and adjust its thermal-comfort threshold accordingly.}

\begin{algorithm}[hbtp!]
\small
\SetAlgoLined
\DontPrintSemicolon
\SetKwInput{Input}{Input}
\SetKwInOut{Output}{Output}
\SetKwInOut{Initialization}{Initialization}
\SetKwComment{Comment}{$\triangleright$\hspace{-4pt}\ }{}\SetInd{0.2em}{0.35em}
\Input{$t_{\rm now}$, \textbf{Initialization:} $\tau \leftarrow 0, {\rm timer} \leftarrow 0$}
\hrule
\If{${\sf occup}_{\rm now}=0\ and\ {\rm timer}=0$}
{
  $T_{\rm target} \leftarrow {\sf setpoint}_{\rm uo}$ \\
  Set HVAC setpoint to $T_{\rm target}$ \Comment*[r]{\textit{increase setpoint}}
 $\tau \leftarrow \emph{Pre-cooling}(t_{\rm now}, t_{\rm pd\_oc})$ \Comment*[r]{\textit{pre-cooling time}}
 ${\rm timer} \leftarrow t_{\rm now} + \tau$
}
\eIf{$t_{\rm now}={\rm timer}$}
{
 $T_{\rm target} \leftarrow {\sf setpoint}_{\rm oc}$ \\
 Set HVAC setpoint to $T_{\rm target}$ \Comment*[r]{\textit{start pre-cooling}}
 ${\rm timer} \leftarrow\ 0$
}
{
  Wait for timer expiry \Comment*[r]{\textit{no pre-cooling yet}}
}
\caption{{\tt HVAC-MPC}}
\label{alg:control}
\end{algorithm}

Having provided an overview of our framework for model predictive control, we will now explain the control algorithm therein. In particular, we call this algorithm {\tt HVAC-MPC}, the pseudocode of which is outlined in Algorithm~\ref{alg:control}. First, in line~1 the algorithm checks whether the building is unoccupied at the current time. If so, then in lines 2 and 3 it instructs the HVAC system to increase the temperature setpoint (see Table~\ref{tab:setpoints} for the exact setpoints used by the algorithm). 

\begin{table}[hbtp!]
  \small
  \caption{Setpoints used by {\tt HVAC-MPC}. Here, ${\sf setpoint}_{\rm oc}$ is used when the building is occupied; ${\sf setpoint}_{\rm uo}$ is used when unoccupied.}
  \label{tab:setpoints}
  \centering
  \begin{tabular}{ c | c}
	\hline \hline
    Occupied (${\sf setpoint}_{\rm oc}$) &  Unoccupied (${\sf setpoint}_{\rm uo}$) \\
    24$^\circ$C & 28$^\circ$C \\
	\hline \hline
   \end{tabular}
\end{table}

After that, in lines 4 and 5, the occupancy prediction is used to determine when the building is expected to become occupied in the future, and determine the optimal pre-cooling time accordingly (here the algorithm uses a procedure called \emph{Pre-cooling}, the workings of which will be explained later on in this section). Then, in line~7, the algorithm checks whether the time has come for the pre-cooling to start. If so, then it instructs the HVAC system to decrease the temperature setpoint, before resetting the corresponding timer (see lines 8 to 10). On the other hand, if the time has not yet come to start pre-cooling, the algorithm simply waits until it does (see lines 11 and 12).

\begin{figure}[hbtp!]
  \scalebox{0.745}{
  \begin{tikzpicture}[node distance = 1.5cm, auto]

    \node (start) [startstop] {Begin};
    \node (init) [init, below of=start, text width=4.5cm, node distance=1.45cm]
          {$t_{\rm min} \gets t_{\rm now}$\\
           $t_{\rm max} \gets t_{\rm pd\_oc}$\\
           $t^* \gets t_{\rm min}+(t_{\rm max}-t_{\rm min})/2$};
    \node (if1) [decision, below of=init, aspect=3.0, node distance=1.85cm]
          {$t_{\rm min}< t_{\rm max}$};
    \node (run) [process, below of=if1, text width=7.3cm, node distance=2.00cm]
          {Run simulation with HVAC turned off from $t_{\rm now}$ 
           to $t^*-1$, and turned on from $t^*$ to $t_{\rm pd\_oc}$};
    \node (if2) [decision, below left = 1.4cm and -1.60cm of run, aspect=2.0]
          {\hspace{0.2em} $T_{\rm sim}[t_{\rm pd\_oc}$+1$]>$\\
           ${\sf setpoint}_{\rm oc}$};
    \node (if3) [decision, right of=if2, aspect=2.0, node distance=5.75cm]
	      {\hspace{0.4em} $T_{\rm sim}[t_{\rm pd\_oc}$--1$]\leq$\\
	       ${\sf setpoint}_{\rm oc}$};
    \node (late) [process, below of=if2, text width=4.9cm, node distance=2.30cm]
          {$t_{\rm max} \gets t^*-1$\\
           $t^* \gets t_{\rm min}+(t^*-t_{\rm min})/2$};
    \node (soon) [process, below of=if3, text width=4.9cm, node distance=2.30cm]
          {$t_{\rm min} \gets t^*+1$\\
           $t^* \gets t^*+(t_{\rm max}-t^*)/2$}; 
    \node (dot) [dot, below left = 0.20cm and 0.30cm of soon] {};
    \node (return) [init, below of=dot, node distance=1.7cm] {Output $t^*$};
    \node (end) [startstop, below of=return, node distance=1.25cm] {End};
    \node (input) [input, right of=run, node distance=4.5cm] {};

    \draw [arrow] (start) -- (init);
    \draw [arrow] (init) -- (if1);
    \draw [arrow] (if1.south) -- node[near start] {yes} (run.north);
    \draw [arrow] (run.south) -- +(0,-11pt) node[align=center, near end, below] {simulated temperature ($T_{\rm sim}$)} -- +(-91pt,-11pt) -- (if2.north);
    \draw [arrow] (input) -- node {\ \ \ \ \ \ \ $T[t_{\rm now}]$} (run.east);
    \draw [arrow] (if2) -- node[align=left] {yes (too late)} (late);
    \draw [arrow] (late.east) -- (dot);
    \draw [arrow] (if2) -- node[align=left] {no} (if3);
    \draw [arrow] (if3) -- node[align=left] {yes (too soon)} (soon);
    \draw [arrow] (soon.west) -- (dot);
    \draw [arrow] (if3.east) -- +(10pt,0) node[near start] {no}  |- +(0,-110pt) -- +(-32pt, -110pt)  -- +(-150pt, -110pt) node[near start, below] {$t^*$ is the best possible pre-cooling time} -- (return.north);
    \draw [arrow] (if1.east) -- +(20pt,0) node[near start] {no} -- +(112pt,0) |- +(0,-267pt) -- (return.east);
    \draw [arrow] (dot.west) -- +(-150pt,0) |- +(-60pt,219pt) -- (if1.west);
    \draw [arrow] (return) -- (end);
  \end{tikzpicture}}
  \caption{Flowchart of our \emph{Pre-cooling} procedure, which determines the pre-cooling time based on the predicted occupancy.}
  \label{fig:eval_precool_flowchart}
\end{figure}
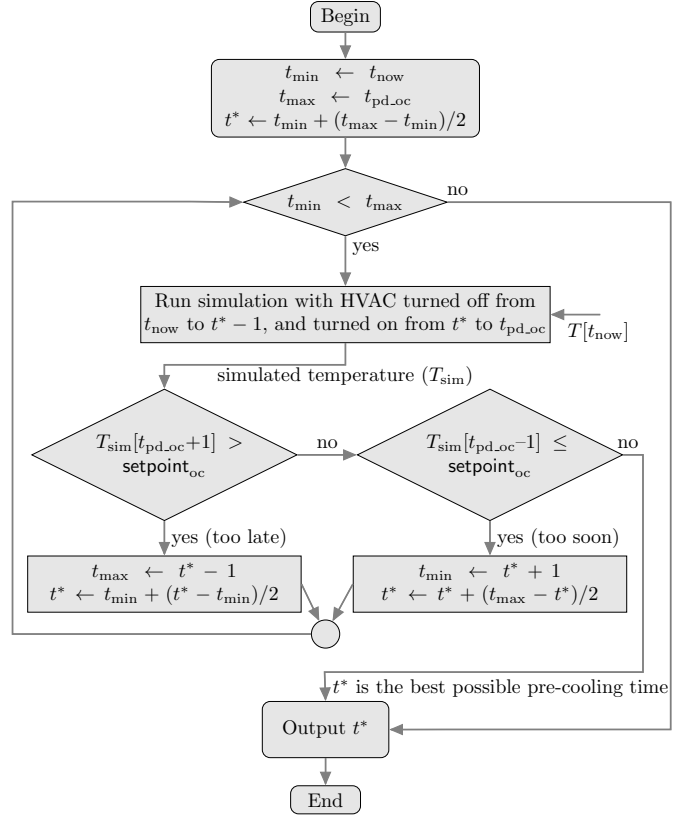

Having explained the pseudocode of {\tt HVAC-MPC}, we now explain the workings of the \emph{Pre-cooling} procedure used therein. Figure~\ref{fig:eval_precool_flowchart} illustrates the flowchart of this procedure. Basically, the goal here is to determine the time $t^*$ at which the pre-cooling should start; this time falls within a certain range of feasible times, denoted by $[t_{\rm min}, t_{\rm max}]$. Initially, the algorithm sets $t_{\rm min}$ to be equal to the current time (i.e., $t_{\rm now}$), sets $t_{\rm max}$ to be equal to the predicted time of occupancy (i.e., $t_{\rm pd\_oc}$), and sets $t^*$ to be right in the middle between the two (i.e., $t^* = t_{\rm min}+(t_{\rm max}-t_{\rm min})/2$). After that, $t^*$ is adjusted iteratively as follows. In each iteration, the algorithm runs an EnergyPlus simulation in which the HVAC is switched off from $t_{\rm min}$ to $t^*-1$, and then the pre-cooling starts at $t^*$, leaving the HVAC on from $t^*$ to $t_{\rm max}$. Thus, we obtain the simulated temperature, $T_{\rm sim}[t]$, for every $t\in [t_{\rm min}, t_{\rm max}]$. These simulated temperatures should ideally satisfy the following conditions:

\begin{figure*}[t]
  \centering
  \begin{subfigure}[b]{0.475\textwidth}
      \centering
      \includegraphics[width=\textwidth]{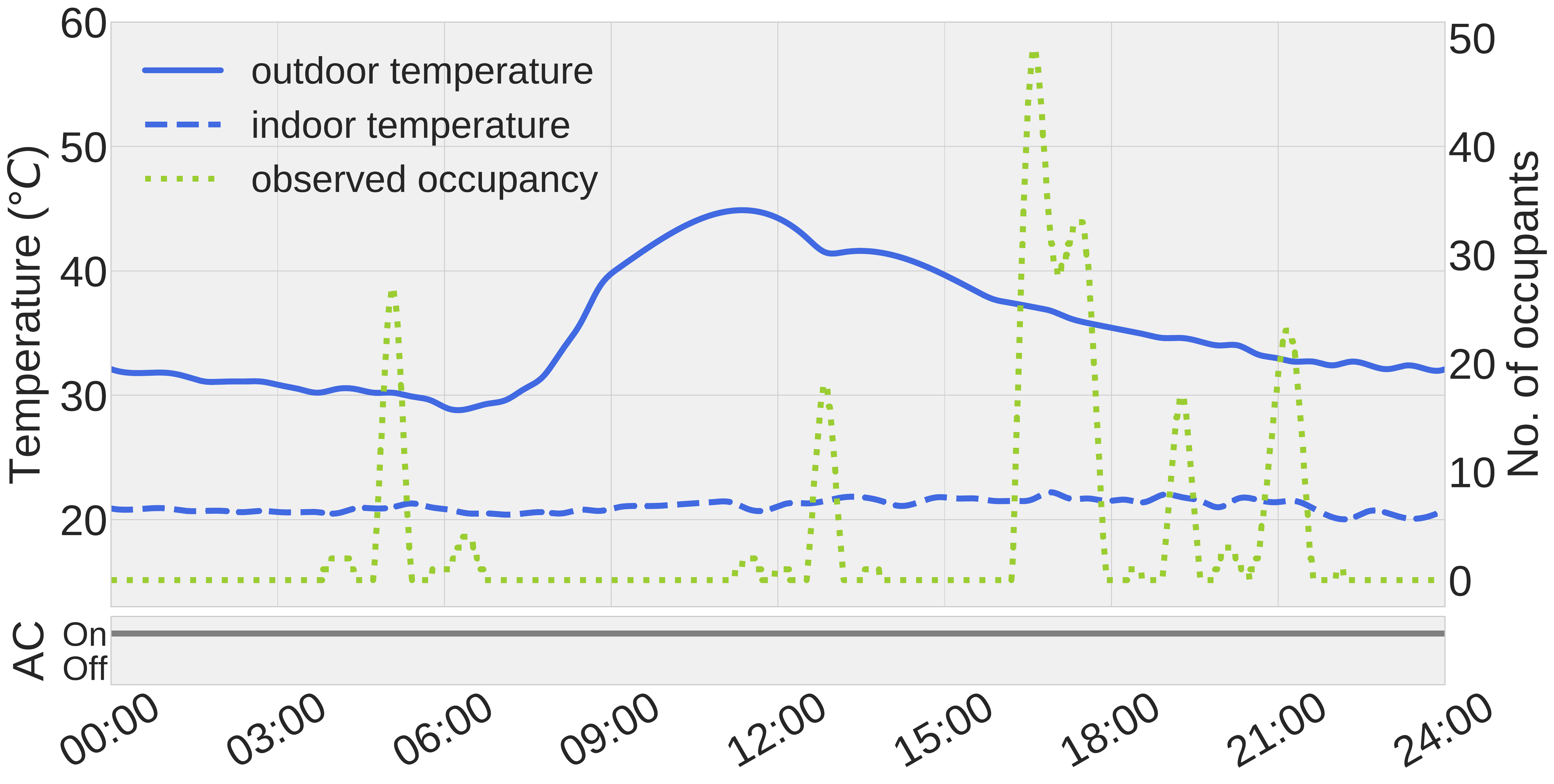}
      \caption{{\bf Without {\tt HVAC-MPC}:} full-powered HVAC at
      all times (indicated by the strip at the bottom) despite
      intermittent occupancy.}
      \label{fig:eval_res_a}
  \end{subfigure}
  \hfill
  \begin{subfigure}[b]{0.475\textwidth}
      \centering
      \includegraphics[width=\textwidth]{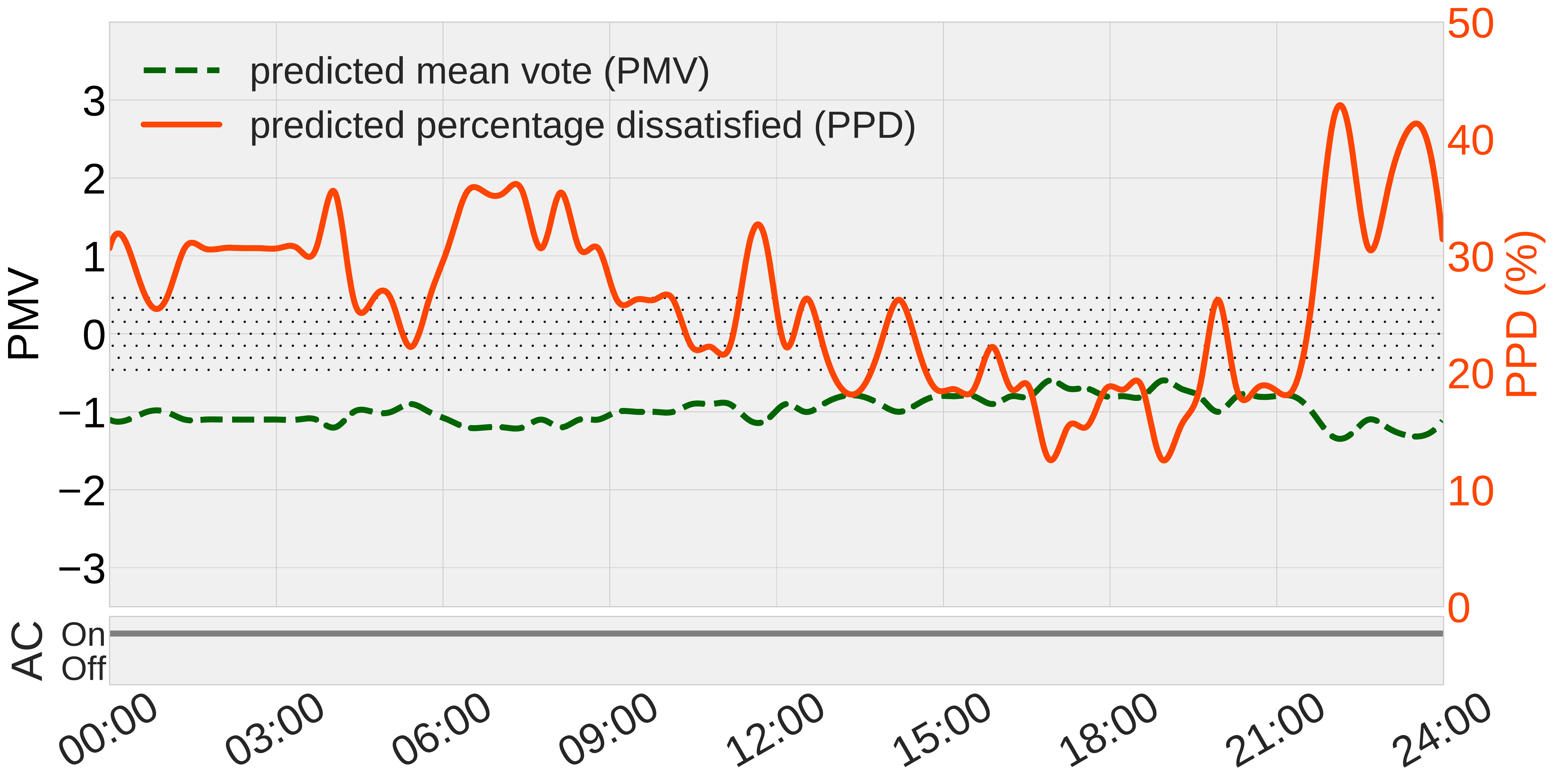}
      \caption{{\bf Without {\tt HVAC-MPC}:} incurred thermal
      comfort sensation. PMV index is always outside the recommended
      [-0.5,+0.5] range.}
      \label{fig:eval_res_b}
  \end{subfigure}
  \begin{subfigure}[b]{0.475\textwidth}
      \centering
      \includegraphics[width=\textwidth]{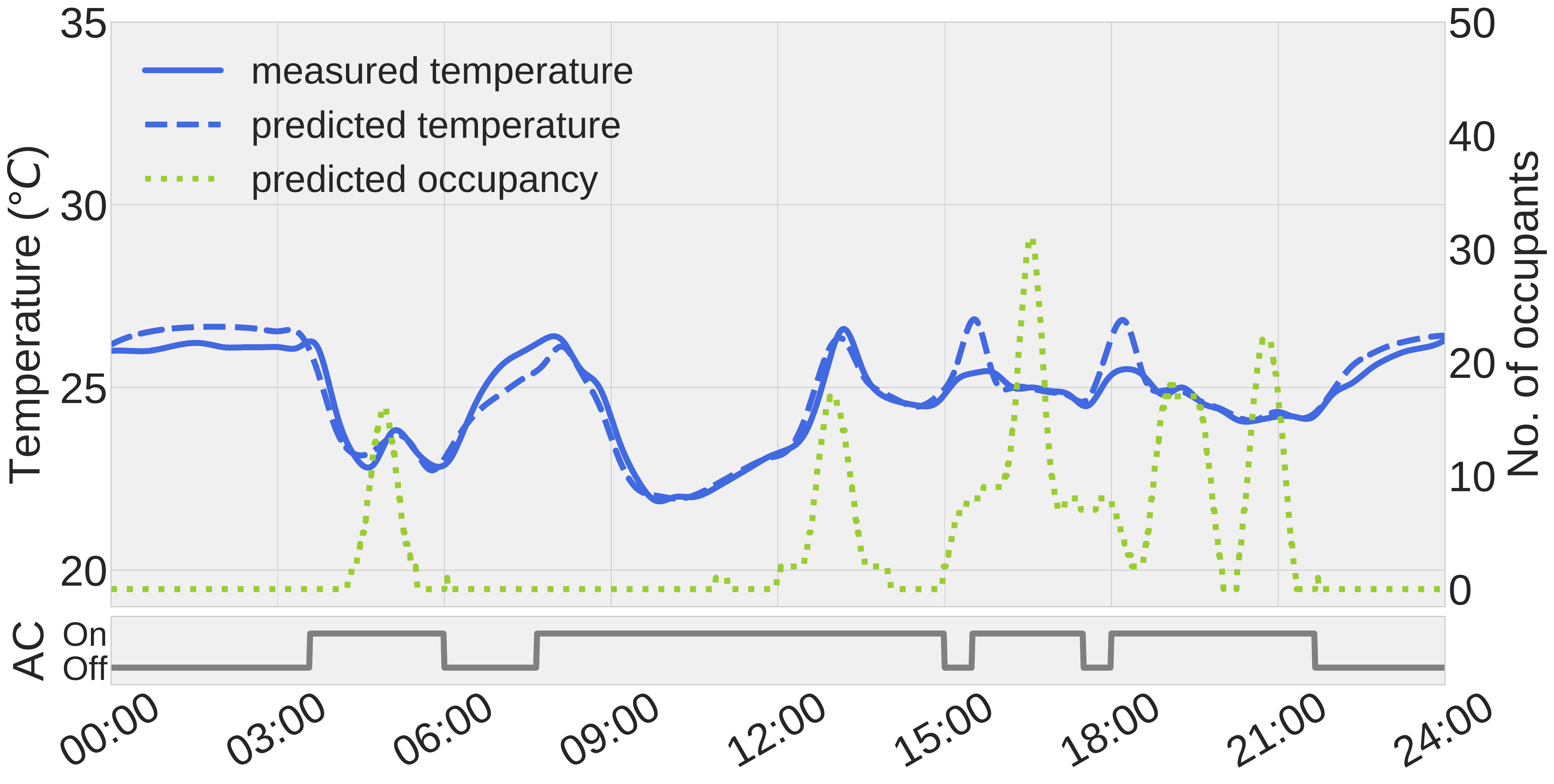}
      \caption{{\bf With {\tt HVAC-MPC}:} occupancy and
      temperature forecasts are used to find the best possible pre-cooling schedule
      as indicated by the sequence of HVAC state transitions from
      {\em off} to {\em on} in the strip at the bottom.}
      \label{fig:eval_res_c}
  \end{subfigure}
  \quad
  \begin{subfigure}[b]{0.475\textwidth}
      \centering
      \includegraphics[width=\textwidth]{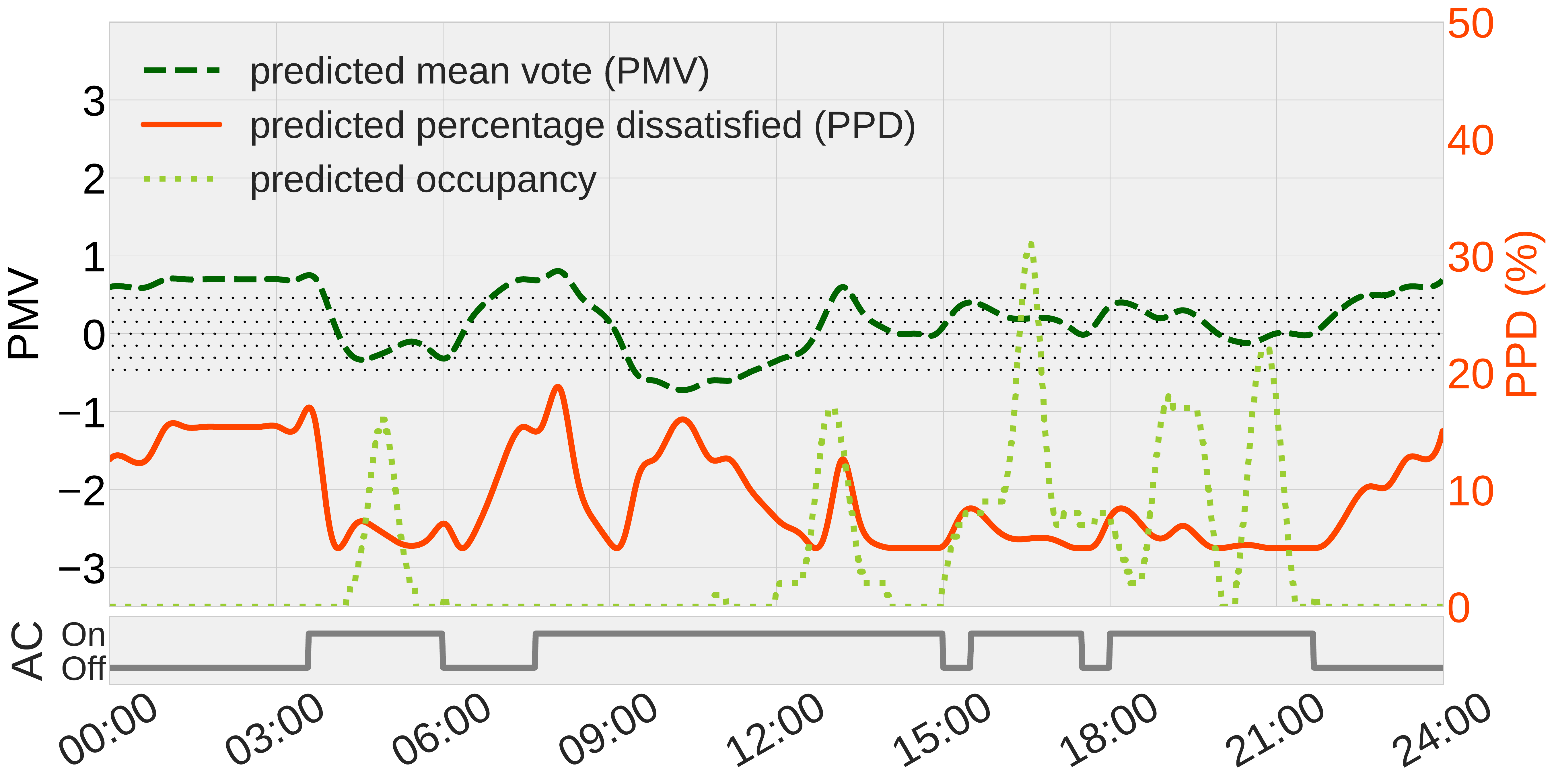}
      \caption{{\bf With {\tt HVAC-MPC}:} The PMV index is almost
      always within the recommended [-0.5,+0.5] range during the periods in which the building is occupied.}
      \label{fig:eval_res_d}
  \end{subfigure}
  \caption[]{Results of HVAC model predictive control ({\tt HVAC-MPC}) in our testbed.}
  \label{fig:eval_results}
\end{figure*}

\begin{equation}\label{condition1}
T_{\rm sim}[t_{\rm pd\_oc}]={\sf setpoint}_{\rm oc}
\end{equation}
\begin{equation}\label{condition2}
T_{\rm sim}[t]>{\sf setpoint}_{\rm oc},\ \forall t : t^* \leq t < t_{\rm pd\_oc}
\end{equation}

\noindent which basically means that the temperature becomes satisfactory exactly when needed, and not before. The algorithm checks whether the above two conditions hold. Now:
\begin{itemize}\itemsep-0.5em
\item if Condition~\eqref{condition1} does not hold, it implies that the pre-cooling process started too late, and so $t^*$ should be set to an earlier time. To this end, the algorithm sets $t_{\rm max}$ to be equal to $t^*-1$, and then updates $t^*$ as follows:
\begin{equation}
t^* \gets t_{\rm min}+(t^*-t_{\rm min})/2
\end{equation}
\item if Condition~\eqref{condition2} does not hold, it implies that the pre-cooling process started too soon, and so $t^*$ should be delayed. To this end, the algorithm sets $t_{\rm min}$ to be equal to $t^*+1$, and then updates $t^*$ as follows:
\begin{equation}
t^* \gets t^*+(t_{\rm max}-t^*)/2
\end{equation}
\end{itemize}
After that, the algorithm proceeds to the next iteration to check whether the new $t^*$ needs any further adjustments. This process is repeated until either $t_{\rm min}\geq t_{\rm max}$, or conditions \eqref{condition1} and \eqref{condition2} are both met. Either way, the best possible $t^*$ is found.


\subsection{Evaluation Results} \label{subsec:eval_res}

\noindent In this subsection, we evaluate the performance of our {\tt HVAC-MPC} algorithm. To this end, we consider two standard measures of thermal comfort, namely \emph{Predicted Mean Vote} (PMV) and \emph{Predicted Percentage Dissatisfied} (PPD) \cite{pmv}. Specifically, PMV quantifies the human perception of thermal sensation on a scale that runs from -3 to +3, where -3 is very cold, 0 is neutral, and +3 is very hot. On the other hand, PPD is built upon PMV to quantify the percentage of occupants that are dissatisfied given the current thermal conditions. {\color{black} The recommended PMV range for thermal comfort is between -0.5 and +0.5 for indoor spaces, while the acceptable PPD range is between 5\% and 10\% \cite{ashrae55}.}

We compared our {\tt HVAC-MPC} algorithm against a baseline alternative, where the HVAC is turned on throughout the day, regardless of the varying occupancy. The evaluation results are shown in Figure~\ref{fig:eval_results}. Specifically:
\begin{itemize}
\item Figure~\ref{fig:eval_res_a} depicts the observed occupancy, the indoor and outdoor temperatures, and the HVAC status \emph{given the baseline HVAC control};
\item Figure~\ref{fig:eval_res_b} depicts the thermal comfort according to PMV and PPD \emph{given the baseline HVAC control};
\item Figure~\ref{fig:eval_res_c} depicts the predicted occupancy, the indoor temperature, and the HVAC status \emph{given {\tt HVAC-MPC} algorithm};
\item Figure~\ref{fig:eval_res_d} depicts the thermal comfort according to PMV and PPD \emph{given {\tt HVAC-MPC} algorithm}.
\end{itemize}

\begin{figure*}[htbp!]
  \begin{subfigure}[b]{0.47\textwidth}
    \centering
    \includegraphics[height=1.6in]{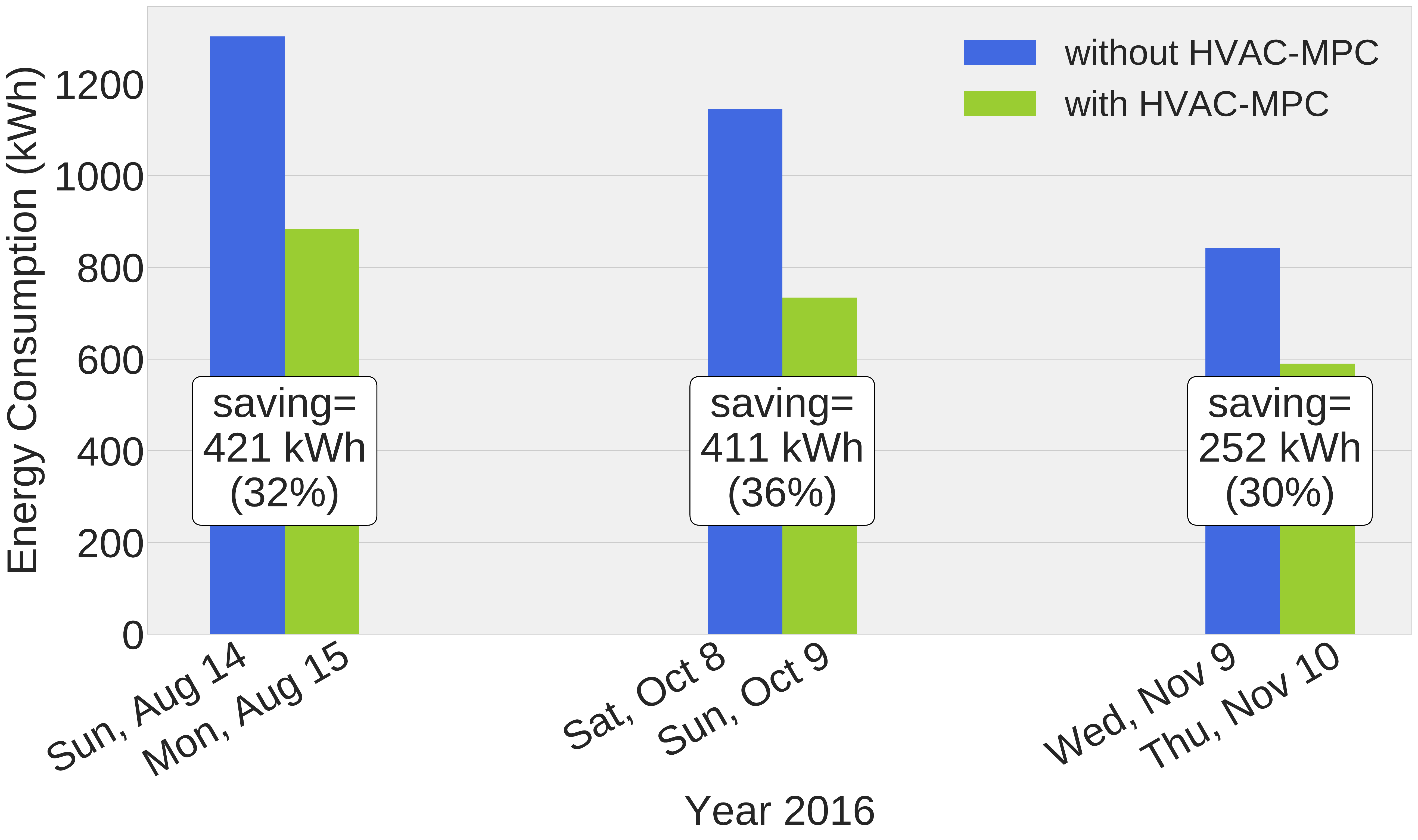}
    \caption{When compared to the immediately preceding day.}
    \label{fig:eval_energy_a}
  \end{subfigure}
  \qquad
  \begin{subfigure}[b]{0.47\textwidth}
    \centering
    \includegraphics[height=1.6in]{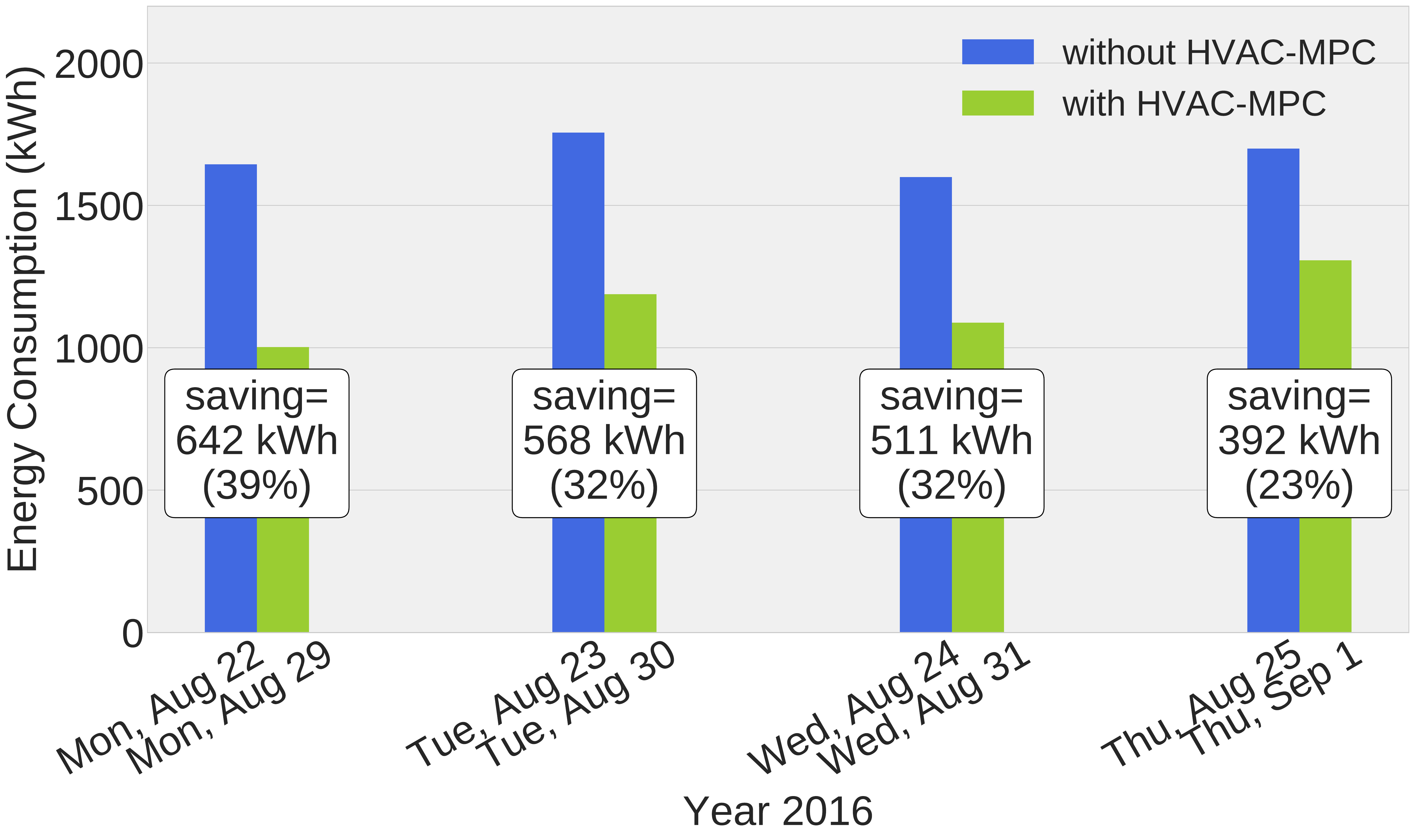}
    \caption{When compared to the same day of the preceding week.}
    \label{fig:eval_energy_b}
  \end{subfigure}
  \caption{Reduction in HVAC energy consumption achieved by {\tt HVAC-MPC} (Algorithm~\ref{alg:control}).}
  \label{fig:eval_energy}
\end{figure*}

Let us first comment on the results of the baseline HVAC control. As can be seen in Figure~\ref{fig:eval_res_a}, the HVAC is turned on throughout the day, despite the significant changes in both the occupancy and the external temperature. In terms of thermal comfort, the baseline HVAC control performs rather poorly, as shown in Figure~\ref{fig:eval_res_b}. Here, the recommended range for PMV is highlighted by the dotted region. As can be seen, the PMV index is outside the recommended range throughout the day. Furthermore, according to the PDD index, a considerable percentage of occupants are dissatisfied most of the day.

Moving on to the results of our {\tt HVAC-MPC} algorithm, Figure~\ref{fig:eval_res_c} shows how the HVAC is switched off for a considerable number of hours. This is because, whenever the building becomes unoccupied, {\tt HVAC-MPC} increases the temperature setpoint and predicts the future temperature and occupancy to find the best possible time for pre-cooling. Figure~\ref{fig:eval_res_d}, on the other hand, depicts the PMV and PPD indices throughout the day. As can be seen, when the building is occupied, the PMV index is almost always within the recommended range. Likewise, when the building is occupied, the PPD is close to 5\% (which is the best possible PPD score that can be achieved).

{\color{black} Finally, we present the energy savings that are attained by our {\tt HVAC-MPC} algorithm. The algorithm was activated in the testbed building for a duration of one week in total and the results are provided in Figure~\ref{fig:eval_energy}. In particular, Figure~\ref{fig:eval_energy_a} shows the results for three different days, where each day is in a different month. For each day shown, the algorithm performance is compared with the immediately preceding day where {\tt HVAC-MPC} is not activated. Figure~\ref{fig:eval_energy_b} shows the results of another experiment where we activated {\tt HVAC-MPC} for four consecutive days in the same week. For each day, the results are compared with the corresponding day of the previous week in which {\tt HVAC-MPC} was not activated. The purpose of this experiment was to find out how much energy is saved by the MPC algorithm relative to the same day of the previous week. The energy savings attained by {\tt HVAC-MPC} range from 23\% to 39\%. Table~\ref{tab:eval_energy} lists the average daily savings, the standard deviation, and the total energy savings over the experiment period.}

\begin{table}[hbtp!]
  \small
  \caption{Summary of energy savings attained by {\tt HVAC-MPC}.}
  \label{tab:eval_energy}
  \centering
  \begin{tabular}{ p{6.5cm} | c}
	\hline \hline
    Average daily savings over experiment period &  456 kWh \\ \hline
    Standard deviation of daily savings &  118 kWh \\ \hline
    Total savings over the seven-day experiment &  3197 kWh \\
	\hline \hline
   \end{tabular}
\end{table}

\section{Testbed Design and Implementation} \label{sec:testbed}

\noindent Our automatic HVAC control system with real-time video-based occupancy recognition and EnergyPlus-guided mod\-el predictive control is implemented on the Raspberry Pi 3 platform. Our choice is motivated by the fact that Raspberry Pi 3 is a low-cost embedded system platform that costs as little as \$35 per unit, as of 2016. Although there are other embedded systems (e.g., Intel Galileo), to the best of our knowledge, Raspberry Pi is the cheapest in the market for our purpose of HVAC control. A comparison of different embedded systems in terms of cost and  performance can be found in \cite{embedded_systems_state_of_the_art}. Another advantage of Raspberry Pi 3 is its rich set of specifications, which include: 1.2GHz 64-bit quad-core CPU, 1GB SDRAM memory, SD card based expandable external data storage, WiFi \& Bluetooth connectivity, and GPU. The remainder of this section provides more details about our testbed design and implementation.

\subsection{System Design}

\begin{figure}[htb!]
 \centering
 \includegraphics[width=\linewidth]{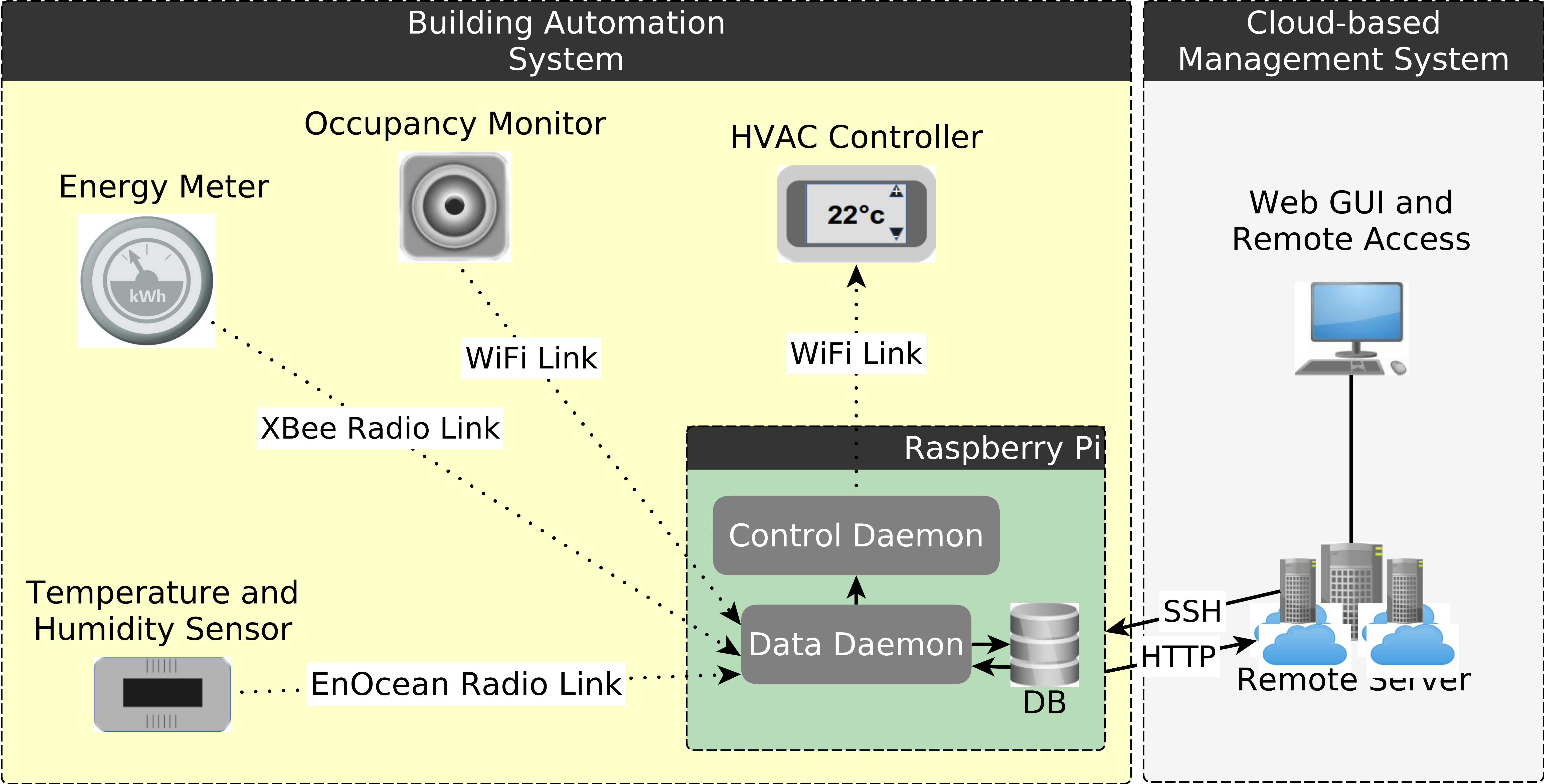}
 \caption{Outline of our system, which is comprised of two subsystems: (1) the building automation system, and (2) the cloud-based management system.}
 \label{fig:system_design}
\end{figure}

\noindent Figure~\ref{fig:system_design} illustrates the overall design of our system. This system is comprised of two subsystems: (1) the building automation system, and (2) the cloud-based management system. Next, we explain each of these systems.

\begin{figure}[htp!]
	\centering
	\begin{subfigure}[b]{0.31\textwidth}
		\centering
		\includegraphics[width=\textwidth]{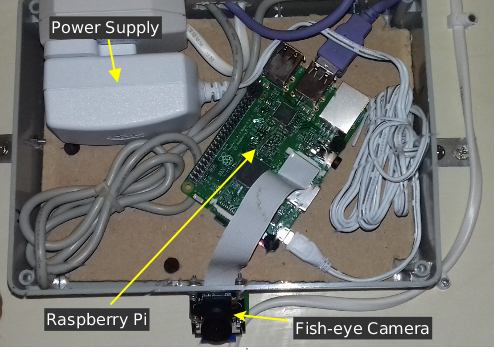}
	     \caption{}
	     \label{fig:occupancy_module}
	\end{subfigure}
	\begin{subfigure}[b]{0.155\textwidth}
		\centering
		\includegraphics[width=\textwidth]{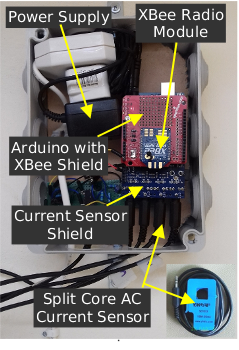}
	     \caption{}
	     \label{fig:energy_module}
	\end{subfigure}
	\begin{subfigure}[b]{0.31\textwidth}
		\centering
		\includegraphics[width=\textwidth]{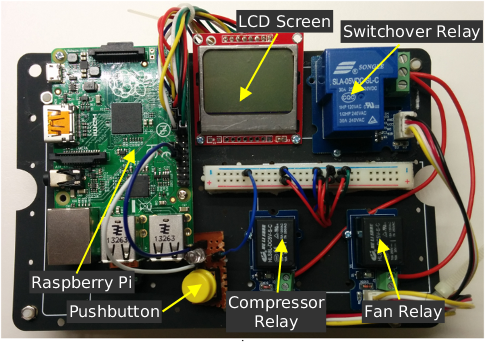}
	     \caption{}
	     \label{fig:hvac_module}
	\end{subfigure}
	\begin{subfigure}[b]{0.153\textwidth}
		\centering
		\includegraphics[width=\textwidth]{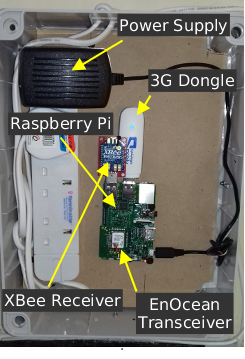}
	    \caption{}
	    \label{fig:base station_module}
	\end{subfigure}
	\caption{Testbed system components: (a) occupancy detection module, (b) HVAC energy measurement module, (c) HVAC controller, (d) base station module.}
	\label{fig:testbed}
\end{figure}

Starting with the building automation system, it consists of the hardware components installed in the building, including temperature and humidity sensors, occupancy monitors, energy meters, and customized wireless HVAC controllers. The temperature and humidity sensors are based on \emph{EnOcean} technology, and have the advantage of being self-powered and maintenance-free, which makes them highly flexible for deployment anywhere in the target environment. The occupancy monitor consists of a dedicated Raspberry Pi and a fish-eye camera for real-time tracking of occupancy. An Arduino-based energy meter is used to measure the energy consumption of the HVAC units. A data daemon running on the base station Raspberry Pi stores the data received from these components. The data daemon also provides the received data to a control daemon (also running on the base station Raspberry Pi), which uses it to make HVAC control decisions. To this end, the control daemon interfaces with the HVAC system through wireless controllers implemented in a Raspberry Pi.  The base station module is shown in Figure \ref{fig:base station_module}.

Having described the building automation system, we now move on to the cloud-based management system. In particular, this system is comprised of the remote server and the web graphical user interface (GUI). The remote server receives the collected data and stores it in a server database for later analysis and permanent storage. The web GUI retrieves data from the server database for visualization and analysis. The web GUI also provides remote access to the building control system. Secure Shell (SSH) protocol is used as the underlying protocol to securely provide remote access.

\subsection{Occupancy Detection Module}

\noindent Our system monitors the building occupancy using a Raspberry Pi  with a fish-eye camera. In order to keep track of the number of occupants in the building, a video processing algorithm runs on the Raspberry Pi to analyze the real-time video stream captured by the fish-eye camera. Figure \ref{fig:occupancy_module} shows the occupancy module developed for this research. The module should be installed above the building entrance door so that it can monitor the occupancy inside the building. Multiple modules may be needed if the building has multiple entry/exit points, such that each module covers a single point.

\subsection{HVAC Energy Measurement Module}

\noindent To collect the HVAC energy consumption data, we developed an Arduino-based module shown in Figure~\ref{fig:energy_module}. The module design is adapted from the \emph{OpenEnergyMonitor} framework, which is an open source project for developing energy monitoring and analysis tools \cite{openenergymonitor}. The module uses non-invasive alternating current (AC) sensors to measure the HVAC energy consumption. These sensors can be clipped onto either the live or ground wire coming into the HVAC unit without needing to strip the wire, thus avoiding any high-voltage work. We used a dedicated sensor for each HVAC unit installed in the building. The collected data is sent to the base station through an XBee radio link. It was decided to use XBee radio technology because the HVAC unit power supply can be in a separate room or even on the roof. In such cases, the XBee modules with their extended transmission range can penetrate concrete walls and roofs and are able to transmit the sensor readings to the base station.

\subsection{HVAC Controller}

\noindent  Figure~\ref{fig:hvac_module} shows the wireless HVAC controller implemented in Raspberry Pi. The module operates the HVAC unit according to the setpoint schedule, which is determined by the MPC algorithm. The module has three relays, two of which are used to open and close the compressor and fan circuits of the HVAC unit to keep the indoor temperature within fixed bound of the MPC setpoint. The third relay (labeled as \emph{Switchover Relay} in Figure~\ref{fig:hvac_module}) is used to switch control between our wireless controller and the default HVAC thermostat controller, allowing the building occupants to disable the automatic HVAC control if they are dissatisfied with the current thermal comfort level or if the system is malfunctioning. The module is also equipped with an LCD display to provide status information. By adjusting the setpoints, we actually control the HVAC system directly. This generic approach can be applied to different HVAC systems, with varying specifications.

\section{Conclusion} \label{sec:concl}

\noindent This paper presents an automatic HVAC control system, featuring real-time occupancy recognition, dynamic occupancy prediction, and simulation-guided model predictive control, implemented in a low-cost embedded system (Raspberry Pi). We deployed and evaluated our system for providing automatic HVAC control in the large public indoor space of a mosque. Our experiments showed that our real-time occupancy recognition system can reach 90\% accuracy, whereas our occupancy prediction system can reach 85\% accuracy. We employ real-time HVAC control guided by an on-board EnergyPlus simulator, which is able to achieve more than 30\% energy saving while maintaining the comfort level within acceptable range. Importantly, our system is sufficiently general and can be deployed in other types of buildings with large public indoor spaces.

Notably, we ported the EnergyPlus simulator to the Raspberry Pi embedded system platform. We release our Raspberry Pi version of EnergyPlus publicly \cite{eplus_rpi} to enable other researchers to take advantage of our work for future building automation projects.

Our testbed has been evaluated in buildings with somewhat predictable occupancy patterns. It is challenging to apply our system to settings with irregular occupancy patterns. In future work, we will explore robust online HVAC control using minimal occupancy prediction. Augmented reality is also being integrated in HVAC control system \cite{ARReprt}. Online algorithms have been applied to wireless sensor based building control \cite{chauToN,AftabeEnergy}. Robust online control can ensure good performance in the presence of dynamic irregular environmental factors.

\section*{Acknowledgments}
\noindent  We would like to thank Afshin Afshari and Prashant Shenoy for helpful discussion.

\bibliography{ref}

\begin{thebibliography}{10}
\expandafter\ifx\csname url\endcsname\relax
  \def\url#1{\texttt{#1}}\fi
\expandafter\ifx\csname urlprefix\endcsname\relax\def\urlprefix{URL }\fi
\expandafter\ifx\csname href\endcsname\relax
  \def\href#1#2{#2} \def\path#1{#1}\fi

\bibitem{HVACConsumption}
L.~Pérez-Lombard, J.~Ortiz, C.~Pout, A review on buildings energy consumption
  information, Energy and Buildings 40~(3) (2008) 394 -- 398.

\bibitem{mpc1}
F.~Oldewurtel, A.~Parisio, C.~N. Jones, D.~Gyalistras, M.~Gwerder, V.~Stauch,
  B.~Lehmann, M.~Morari, Use of model predictive control and weather forecasts
  for energy efficient building climate control, Energy and Buildings 45 (2012)
  15 -- 27.

\bibitem{mpc2}
J.~{\'A}lvarez, J.~Redondo, E.~Camponogara, J.~Normey-Rico, M.~Berenguel,
  P.~Ortigosa, Optimizing building comfort temperature regulation via model
  predictive control, Energy and Buildings 57 (2013) 361--372.

\bibitem{mpc3}
S.~Salakij, N.~Yu, S.~Paolucci, P.~Antsaklis, Model-based predictive control
  for building energy management. i: Energy modeling and optimal control,
  Energy and Buildings 133 (2016) 345 -- 358.

\bibitem{mpc5}
Y.~Kwak, J.-H. Huh, C.~Jang, Development of a model predictive control
  framework through real-time building energy management system data, Applied
  Energy 155 (2015) 1 -- 13.

\bibitem{mpc6}
P.-D. Moro{\c{s}}an, R.~Bourdais, D.~Dumur, J.~Buisson, Building temperature
  regulation using a distributed model predictive control, Energy and Buildings
  42~(9) (2010) 1445--1452.

\bibitem{mpc7}
F.~Ascione, N.~Bianco, C.~D. Stasio, G.~M. Mauro, G.~P. Vanoli,
  Simulation-based model predictive control by the multi-objective optimization
  of building energy performance and thermal comfort, Energy and Buildings 111
  (2016) 131 -- 144.

\bibitem{lti_rc}
H.~Park, M.~Ruellan, A.~Bouvet, E.~Monmasson, R.~Bennacer, Thermal parameter
  identification of simplified building model with electric appliance, in: 11th
  International Conference on Electrical Power Quality and Utilisation, 2011,
  pp. 1--6.

\bibitem{mpc_rc_calib1}
R.~B{\u{a}}lan, J.~Cooper, K.-M. Chao, S.~Stan, R.~Donca, Parameter
  identification and model based predictive control of temperature inside a
  house, Energy and Buildings 43~(2) (2011) 748--758.

\bibitem{mpc_rc_calib2}
J.~Hu, P.~Karava, A state-space modeling approach and multi-level optimization
  algorithm for predictive control of multi-zone buildings with mixed-mode
  cooling, Building and Environment 80 (2014) 259 -- 273.

\bibitem{soc}
C.-H. Jan, 10 years of transistor innovations in system-on-chip (soc) era, in:
  Solid-State and Integrated Circuit Technology (ICSICT), 2014 12th IEEE
  International Conference on, IEEE, 2014, pp. 1--4.

\bibitem{rpi}
Raspberry pi foundation, available at: \url{https://www.raspberrypi.org}.
  Retrieved on 31-Jan-2017.

\bibitem{embedded_ref2}
B.~Pavlin, G.~Pernigotto, F.~Cappelletti, P.~Bison, R.~Vidoni, A.~Gasparella,
  Real-time monitoring of occupants' thermal comfort through infrared imaging:
  A preliminary study, Buildings 7~(1) (2017) 10.

\bibitem{eplus}
D.~B. Crawley, L.~K. Lawrie, F.~C. Winkelmann, W.~Buhl, Y.~Huang, C.~O.
  Pedersen, R.~K. Strand, R.~J. Liesen, D.~E. Fisher, M.~J. Witte, J.~Glazer,
  Energyplus: creating a new-generation building energy simulation program,
  Energy and Buildings 33~(4) (2001) 319 -- 331, special Issue: \{BUILDING\}
  SIMULATION'99.

\bibitem{eplus_rpi}
Energyplus for raspberry pi, available at:
  \url{https://github.com/muhaftab/energyplus_rpi}. Retrieved on 05-May-2017.

\bibitem{NumberOfMusquesInUAE}
Number of mosques in u.a.e, available at:
  \url{https://www.awqaf.gov.ae/Affair.aspx?SectionID=3&RefID=18}. Retrieved on
  20-Jan-2017.

\bibitem{NumberOfMusquesInSaudi}
Number of mosques in k.s.a, available at:
  \url{http://www.moia.gov.sa/menu/pages/statistics.aspx}. Retrieved on
  20-Jan-2017.

\bibitem{pir1}
D.~T. Delaney, G.~M. O'Hare, A.~G. Ruzzelli, Evaluation of energy-efficiency in
  lighting systems using sensor networks, in: Proceedings of ACM Workshop on
  Embedded Sensing Systems for Energy-Efficiency in Buildings, ACM, 2009, pp.
  61--66.

\bibitem{pir2}
A.~Barbato, L.~Borsani, A.~Capone, S.~Melzi, Home energy saving through a user
  profiling system based on wireless sensors, in: Proceedings of ACM workshop
  on embedded sensing systems for energy-efficiency in buildings, ACM, 2009,
  pp. 49--54.

\bibitem{pir3}
R.~S. Hsiao, D.~B. Lin, H.~P. Lin, S.~C. Cheng, C.~H. Chung, A robust
  occupancy-based building lighting framework using wireless sensor networks,
  in: Applied Mechanics and Materials, Vol. 284, Trans Tech Publ, 2013, pp.
  2015--2020.

\bibitem{camera1}
J.~Li, L.~Huang, C.~Liu, Robust people counting in video surveillance: Dataset
  and system, in: Advanced Video and Signal-Based Surveillance (AVSS), 2011 8th
  IEEE International Conference on, IEEE, 2011, pp. 54--59.

\bibitem{camera2}
L.~Chen, F.~Chen, X.~Guan, A video-based indoor occupant detection and
  localization algorithm for smart buildings, in: Emerging Intelligent
  Computing Technology and Applications, Springer, 2009, pp. 565--573.

\bibitem{review_pred1}
M.~Jia, R.~S. Srinivasan, Occupant behavior modeling for smart buildings: A
  critical review of data acquisition technologies and modeling methodologies,
  in: Winter Simulation Conference (WSC), 2015, IEEE, 2015, pp. 3345--3355.

\bibitem{review_pred2}
W.~Kleiminger, F.~Mattern, S.~Santini, Predicting household occupancy for smart
  heating control: A comparative performance analysis of state-of-the-art
  approaches, Energy and Buildings 85 (2014) 493--505.

\bibitem{dong2011-1}
B.~Dong, K.~P. Lam, C.~Neuman, Integrated building control based on occupant
  behavior pattern detection and local weather forecasting, in: Twelfth
  International IBPSA Conference. Sydney: IBPSA Australia, Citeseer, 2011, pp.
  14--17.

\bibitem{dong2014}
B.~Dong, K.~P. Lam, A real-time model predictive control for building heating
  and cooling systems based on the occupancy behavior pattern detection and
  local weather forecasting, in: Building Simulation, Vol.~7, Springer, 2014,
  pp. 89--106.

\bibitem{neuralnetworks}
P.~Ferreira, A.~Ruano, S.~Silva, E.~Conceição, Neural networks based
  predictive control for thermal comfort and energy savings in public
  buildings, Energy and Buildings 55 (2012) 238 -- 251, cool Roofs, Cool
  Pavements, Cool Cities, and Cool World.

\bibitem{narmax}
Identification of nonlinear systems using narmax model, Nonlinear Analysis:
  Theory, Methods \& Applications 71~(12) (2009) e1198 -- e1202.

\bibitem{simulators}
D.~B. Crawley, J.~W. Hand, M.~Kummert, B.~T. Griffith, Contrasting the
  capabilities of building energy performance simulation programs, Building and
  Environment 43~(4) (2008) 661 -- 673, part Special: Building Performance
  Simulation.

\bibitem{calib_review}
D.~Coakley, P.~Raftery, M.~Keane, A review of methods to match building energy
  simulation models to measured data, Renewable and Sustainable Energy Reviews
  37 (2014) 123 -- 141.

\bibitem{cosim_mpc4}
J.~Cigler, D.~Gyalistras, J.~{\v{S}}iroky, V.~Tiet, L.~Ferkl, Beyond theory:
  the challenge of implementing model predictive control in buildings, in:
  Proceedings of 11th Rehva World Congress, Clima, Vol. 250, 2013.

\bibitem{sim_based_mpc}
J.~Zhao, K.~P. Lam, B.~E. Ydstie, V.~Loftness, Occupant-oriented mixed-mode
  energyplus predictive control simulation, Energy and Buildings 117 (2016) 362
  -- 371.

\bibitem{cosim_mpc5}
D.~Sturzenegger, D.~Gyalistras, M.~Gwerder, C.~Sagerschnig, M.~Morari, R.~S.
  Smith, Model predictive control of a swiss office building, in: Clima-rheva
  world congress, 2013, pp. 3227--3236.

\bibitem{dong2009}
B.~Dong, B.~Andrews, Sensor-based occupancy behavioral pattern recognition for
  energy and comfort management in intelligent buildings, in: Proceedings of
  building simulation, 2009, pp. 1444--1451.

\bibitem{dong2011-2}
B.~Dong, K.~P. Lam, Building energy and comfort management through occupant
  behaviour pattern detection based on a large-scale environmental sensor
  network, Journal of Building Performance Simulation 4~(4) (2011) 359--369.

\bibitem{goyal2013}
S.~Goyal, H.~A. Ingley, P.~Barooah, Occupancy-based zone-climate control for
  energy-efficient buildings: Complexity vs. performance, Applied Energy 106
  (2013) 209--221.

\bibitem{opencv_library}
G.~Bradski, Dr. Dobb's Journal of Software Tools.

\bibitem{ox_library}
openFrameworks Community, openframeworks.

\bibitem{zivkovic2006efficient}
Z.~Zivkovic, F.~van~der Heijden, Efficient adaptive density estimation per
  image pixel for the task of background subtraction, Pattern recognition
  letters 27~(7) (2006) 773--780.

\bibitem{zivkovic2004improved}
Z.~Zivkovic, Improved adaptive gaussian mixture model for background
  subtraction, in: Proceedings of IEEE International Conference on Pattern
  Recognition, Vol.~2, IEEE, 2004, pp. 28--31.

\bibitem{Suzuki85Topological}
S.~Suzuki, K.~Abe, Topological structural analysis of digitized binary images
  by border following, CVGIP 30~(1) (1985) 32--46.

\bibitem{halko2011finding}
N.~Halko, P.-G. Martinsson, J.~A. Tropp, Finding structure with randomness:
  Probabilistic algorithms for constructing approximate matrix decompositions,
  SIAM review 53~(2) (2011) 217--288.

\bibitem{scikit-learn}
F.~Pedregosa, G.~Varoquaux, A.~Gramfort, V.~Michel, B.~Thirion, O.~Grisel,
  M.~Blondel, P.~Prettenhofer, R.~Weiss, V.~Dubourg, J.~Vanderplas, A.~Passos,
  D.~Cournapeau, M.~Brucher, M.~Perrot, E.~Duchesnay, Scikit-learn: Machine
  learning in {P}ython, Journal of Machine Learning Research 12 (2011)
  2825--2830.

\bibitem{f1score1}
Y.~Yang, X.~Liu, A re-examination of text categorization methods, in:
  Proceedings of the 22nd annual international ACM SIGIR conference on Research
  and development in information retrieval, ACM, 1999, pp. 42--49.

\bibitem{f1score2}
W.~Cohen, P.~Ravikumar, S.~Fienberg, A comparison of string metrics for
  matching names and records, in: Kdd workshop on data cleaning and object
  consolidation, Vol.~3, 2003, pp. 73--78.

\bibitem{eplus_weather}
Energyplus standard weather data, available at:
  \url{https://energyplus.net/weather}. Retrieved on 08-Feb-2017.

\bibitem{bcvtb}
M.~Wetter, Co-simulation of building energy and control systems with the
  building controls virtual test bed, Journal of Building Performance
  Simulation 4~(3) (2011) 185--203.

\bibitem{cvrmse_mbe}
M.~Royapoor, T.~Roskilly, Building model calibration using energy and
  environmental data, Energy and Buildings 94 (2015) 109 -- 120.

\bibitem{idd}
Energyplus input data dictionary (idd), available at:
  \url{https://energyplus.net/sites/default/files/pdfs/pdfs_v8.3.0/InputOutputReference.pdf}.
  Retrieved on 05-May-2017.

\bibitem{ashrae14}
A.~Guideline, Ansi/ashrae standard 14-2014, Measurement of energy, demand, and
  water savings.

\bibitem{pmv}
P.~O. Fanger, et~al., Thermal comfort. analysis and applications in
  environmental engineering., Thermal comfort. Analysis and applications in
  environmental engineering.

\bibitem{ashrae55}
A.~Standard, Ansi/ashrae standard 55-2004, Thermal Environmental Conditions for
  Human Occupancy.

\bibitem{embedded_systems_state_of_the_art}
Embedded linux board comparison, available at:
  \url{https://learn.adafruit.com/embedded-linux-board-comparison?view=all}.
  Retrieved on 05-May-2017.

\bibitem{openenergymonitor}
Open energy monitor -- a project to develop and build open source energy
  monitoring and analysis tools, available at:
  \url{https://openenergymonitor.org/}. Retrieved on 07-Feb-2017.

\bibitem{ARReprt}
M.~Aftab, C.-K. Chau, Enabling self-aware smart buildings by augmented reality,
  Tech. rep. (2017).

\bibitem{chauToN}
C.-K. Chau, M.~Khonji, M.~Aftab, Online algorithms for information aggregation
  from distributed and correlated sources, IEEE/ACM Transactions on Networking
  24~(6) (2016) 3714--3725.

\bibitem{AftabeEnergy}
M.~Aftab, C.-K. Chau, P.~Armstrong, Smart air-conditioning control by wireless
  sensors: An online optimization approach, in: Proceedings of ACM
  International Conference on Future Energy Systems (e-Energy), 2013, pp.
  225--236.

\end{thebibliography}

\end{document}